\documentclass[conference]{IEEEtran}

\usepackage{graphicx}
\usepackage{lipsum}
\usepackage{hyperref}
\usepackage{cite}
\usepackage{pifont}
\usepackage{enumitem}
\usepackage{caption}
\usepackage{subcaption}
\usepackage{mathabx}
\usepackage{amsmath}
\usepackage{adjustbox}
\newtheorem{theorem}{Theorem}
\IEEEoverridecommandlockouts
\usepackage{atbegshi}
\AtBeginDocument{\AtBeginShipoutNext{\AtBeginShipoutDiscard}}

\begin{document}

\title{\LARGE Joint Hybrid Transceiver and Reflection Matrix Design for RIS-Aided mmWave MIMO Cognitive Radio Systems}
\author{\small
Jitendra~Singh, \textit{Student Member,~IEEE,}
Suraj~Srivastava,  \textit{Senior Member,~IEEE,}
Surya~P.~Yadav, \textit{Student Member,~IEEE,}\\
Aditya~K.~Jagannatham, \textit{Senior Member,~IEEE,}
and Lajos~Hanzo, \textit{Life Fellow,~IEEE}\vspace{-45pt}
}
\thanks{J. Singh, S. P. Yadav, and A. K. Jagannatham are with the Department of Electrical Engineering, Indian Institute of Technology Kanpur, Kanpur, UP 208016, India (e-mail: jitend@iitk.ac.in; sprakashy21@iitk.ac.in; adityaj@iitk.ac.in).}
\thanks{S. Srivastava is with the Department of Electrical Engineering, Indian Institute of Technology Jodhpur, Jodhpur, Rajasthan 342030, India (email: surajsri@iitj.ac.in).}
\thanks{L. Hanzo is with the School of Electronics and Computer Science, University of Southampton, Southampton SO17 1BJ, U.K. (e-mail: lh@ecs.soton.ac.uk).}

\maketitle
\begin{abstract}
In this work, a reconfigurable intelligent surface (RIS)-aided millimeter wave (mmWave) multiple-input multiple-output (MIMO) cognitive radio (CR) downlink operating in the underlay mode is investigated. The cognitive base station (CBS) communicates with multiple secondary users (SUs), each having multiple RF chains in the presence of a primary user (PU). We conceive a joint hybrid transmit precoder (TPC), receiver combiner (RC), and RIS reflection matrix (RM) design, which maximizes the sum spectral efficiency (SE) of the secondary system while maintaining the interference induced at the PU below a specified threshold. To this end, we formulate the sum-SE maximization problem considering the total transmit power (TP), the interference power (IP), and the non-convex unity modulus constraints of the RF TPC, RF RC, and RM. To solve this highly non-convex problem, we propose a two-stage hybrid transceiver design in conjunction with a novel block coordinate descent (BCD)-successive Riemannian conjugate gradient (SRCG) algorithm. 
We initially decompose the RF TPC, RC, and RM optimization problem into a series of sub-problems and subsequently design pairs of RF TPC and RC vectors, followed by successively optimizing the elements of the RM using the iterative BCD-SRCG algorithm. 
Furthermore, based on the effective baseband (BB) channel, the BB TPC and BB RC are designed using the proposed direct singular value decomposition (D-SVD) and projection based SVD (P-SVD) methods. Subsequently, the proportional water-filling solution is proposed for optimizing the power, which maximizes the weighted sum-SE of the system. Finally, simulation results are provided to compare our proposed schemes to several benchmarks and quantify the impact of other parameters on the sum-SE of the system. 
\end{abstract}



\maketitle

\section{\uppercase{INTRODUCTION}}
\IEEEPARstart{T}{he} growing need for high data rates has spurred the development of new technologies, such as 6G wireless communication networks. Given the high bandwidth requirements of these networks, the mmWave band spaning the frequency band of 30-300 GHz is eminently suitable for next-generation wireless systems \cite{HBF_17,mm1}. However, in comparison to the conventional sub-6 GHz bands, the mmWave band suffers from severe path, loss as well as penetration, and absorption losses \cite{HBF_1,HBF_2}. To overcome these losses, high gain multiple-input-multiple-output (MIMO) schemes have been recommended for mmWave systems.

Although the resultant large antenna arrays are suitable for mmWave MIMO systems, the signals in the high frequency regime are highly susceptible to blockages, which can adversely affect mmWave MIMO systems. In this context, reconfigurable intelligent surface (RIS) 
technology \cite{IRS_mm_16,IRS_mm_13,IRS_mm_14,IRS_16,IRS_mm_15} can play a crucial role by providing an alternative path for communication. RISs are made of low-cost reflecting elements and power efficient passive beamforming can be carried out by harnessing them. 

Furthermore, due to the highly directive nature of mmWave technology, the interference between different wireless networks operating in the same frequency band is reduced due to their highly directional beams \cite{CR_A1,CR_A2,CR_A3}. This presents an excellent opportunity for mmWave MIMO systems to be used in cognitive radio (CR) systems. In such systems \cite{CR_3,HBF_16}, the secondary users (SUs) opportunistically harness the same frequency band as the primary users (PUs) without significantly affecting the PU's communication. This motivates us to study the impact of RIS on mmWave MIMO CR systems, which can potentially maximize the efficiency of RIS-aided mmWave MIMO systems. Many researchers have studied the active transmit precoder (TPC) and passive RIS reflection matrix (RM) design in RIS-aided mmWave MIMO systems. The related literature survey is discussed in the next subsection.
\subsection{Literature review} \label{literature review}
CR is a revolutionary technology that provides high spectrum utilization for wireless communications by allowing the SUs to access the radio spectrum or share the unused spectrum of PUs without degrading the quality of service of the PUs \cite{CR_3}. The authors of \cite{CR_3} have designed an underlay spectrum sharing scheme for CR networks, where the SUs communicate with the cognitive base station (CBS) with controlled power, which does not affect the quality of service of the PUs. Moreover, the authors of \cite{IRS_CR_1,IRS_CR_2,IRS_CR_3,IRS_CR_10,IRS_CR_4,IRS_CR_5,IRS_CR_6,IRS_CR_7,IRS_CR_8,IRS_CR_9,IRS_CR_11} consider RIS-aided MIMO CR systems. Specifically, Tian \textit{et al.} \cite{IRS_CR_1} summarized the potential of RIS-aided spectrum sharing systems and discussed the diverse practical use cases of these systems in vehicular and UAV communication. The authors of \cite{IRS_CR_2} proposed joint active beamforming at the CBS and passive beamforming at the RIS for maximizing the weighted sum-rate of the SUs under total transmit power (TP) and interference power (IP) constraints in an RIS-aided MIMO CR system. They have used a block coordinate descent (BCD) algorithm to design the active TPC and passive beamforming based on full channel state information (CSI). Furthermore, to incorporate imperfect CSI in RIS-aided MIMO CR systems, Zhang \textit{et al.} \cite{IRS_CR_3} proposed joint active and passive beamforming for minimizing total TP at the CBS.
Jiang \textit{et al.}\cite{IRS_CR_10} consider an underlay RIS-aided MIMO CR system and proposed joint active and passive beamformer designs for maximizing the weighted sum-SE of the SUs under specific TP and IP constraints. They reformulate the resultant non-convex problem as a pair of sub-optimization problems using the weighted minimum mean-square error (WMMSE) criterion and subsequently optimize the TPC and RIS RM using the popular alternating optimization method.  
\begin{table*}[t!]
    \centering
    \caption{Contrasting our novel contributions to the literature of RIS-aided mmWave MIMO CR systems} \label{tab:lit_rev}
\begin{tabular}{|l|c|c|c|c|c|c|c|c|c|c|c|c|c|c|c|c|}
    \hline
 &\cite{CR_A2}  &\cite{CR_3}   &\cite{IRS_CR_2}   &\cite{IRS_CR_3}  &\cite{IRS_CR_10} &\cite{HBF_3_1}  &\cite{HBF_3_2}  &\cite{HBF_4}  &\cite{HBF_13}  &\cite{HBF_14} &\cite{IRS_mm_6}  &\cite{IRS_mm_3}  &\cite{IRS_mm_4}  & \cite{IRS_mm_9} & Proposed \\ [0.5ex]
 \hline
Underlay CR  &\checkmark  &\checkmark  &\checkmark  &\checkmark  &\checkmark  &  &  &  &\checkmark  &\checkmark  &   &  &   &  &\checkmark\\
\hline
mmWave MIMO  &\checkmark  &  &  &  &  &\checkmark &\checkmark  &\checkmark  &\checkmark  &\checkmark &\checkmark  &\checkmark &\checkmark  &\checkmark & \checkmark\\
\hline
RIS   &  &  &\checkmark  &\checkmark  &\checkmark  &  &  &  &  &  &\checkmark  &\checkmark  &\checkmark  &\checkmark &\checkmark\\
\hline
Multi-user &\checkmark  &\checkmark  &\checkmark  &\checkmark   &\checkmark &\checkmark &\checkmark  &\checkmark  &\checkmark  &\checkmark  &   &\checkmark  &\checkmark  &\checkmark &\checkmark\\
 \hline
Multiple RF chain per user
 &\checkmark  &   &\checkmark  &  &\checkmark  &  &  &\checkmark  &\checkmark  &\checkmark &\checkmark  &   &\checkmark  & \checkmark & \checkmark\\
 \hline
Hybrid TPC  &\checkmark  &  &  &  &  &\checkmark &\checkmark  &\checkmark  &\checkmark &\checkmark &\checkmark  &\checkmark  &\checkmark &\checkmark  &\checkmark\\
 \hline
Joint active and passive beamforming  &  &  &\checkmark &\checkmark  &\checkmark  &  &  &  & &  &\checkmark  &\checkmark  &\checkmark  &\checkmark &\checkmark\\
 \hline
Sum-SE maximization &\checkmark  &\checkmark  &\checkmark  &  &\checkmark &\checkmark &\checkmark  &\checkmark &\checkmark  &\checkmark  &\checkmark &   &  & \checkmark &\checkmark\\
 \hline
BCD method &  &  &\checkmark &  &  &  &  &  &  &  &  &\checkmark &  &\checkmark &\checkmark\\
 \hline
Two-stage hybrid TPC
&   &  &  &  &  &  &\checkmark  &\checkmark  &  &\checkmark  &   &  &  &  &\checkmark \\
 \hline
Zero-forcing  &  &\checkmark &  &  &  &  &\checkmark  &\checkmark  &  &\checkmark  &   &  &\checkmark  &  &\checkmark\\
 \hline
Optimal power allocation  &\checkmark  &\checkmark  &  &  &  &  &  &  &  &\checkmark &\checkmark  &  &\checkmark &  &\checkmark\\
 \hline
Direct-SVD &  &\checkmark  &  &\checkmark   &\checkmark   &  &  &  &  &\checkmark &  &  & &  &\checkmark\\
 \hline
{\bf Projection based-SVD} &  &  &  &  &  &  &  &  &  & &  &  & &  &\checkmark\\
 \hline
{\bf Successive RCG optimization} &  &  &  &  &  &  &  & &  &  &  &   &  &  &\checkmark\\
 \hline 
{\bf Proportional water filling}  &  &  &  &  &  &  &  &  &  & &  &  & &  &{\bf \checkmark}\\
 \hline
\end{tabular}
\end{table*}
Lin \textit{et al.} \cite{IRS_CR_4} consider an RIS for spectrum sensing in CR systems, where they proposed a weighted energy detection method operating in the presence of a PU in RIS aided CR networks. Zamanian \textit{et al.} \cite{IRS_CR_5} incorporate a vertical beamforming mechanism at the CBS for maximizing the SE in an RIS-aided CR network, where they jointly optimize the active and passive beamformer. They concluded that the SE of the system is maximized, when the tilt angles at the CBS are oriented towards the RIS. Moreover, Dong \textit{et al.} \cite{IRS_CR_6} maximize the secrecy rate of the SUs in an RIS-aided multiple input single output (MISO) wiretap channel operating in the underlay mode both under perfect and imperfect CSI of the eavesdropper. Furthermore, Zhang \textit{et al.} \cite{IRS_CR_7} consider a symbol level precoder at the CBS to minimize the symbol error rate of an RIS-aided MISO CR system. They use the alternating optimization technique for designing the active and passive beamformers, where a successive convex approximation (SCA) method is adopted for optimizing the RIS RM. Yang \textit{et al.} \cite{IRS_CR_8} analyze the outage probability in an RIS-aided CR system, where an RIS has been used for eliminating the interference at the PU caused by the SU. Furthermore, Vu \textit{et al.} \cite{IRS_CR_9} investigated an underlay RIS-aided non-orthogonal multiple access (NOMA) CR network, wherein they derived the outage probability of the SUs, the sum throughput and ergodic capacity of the system considering both line of sight (LoS) and non-LoS communication (NLoS) links between the CBS and SUs. 

However, the authors of \cite{IRS_CR_1,IRS_CR_2,IRS_CR_3,IRS_CR_4,IRS_CR_5,IRS_CR_6,IRS_CR_7,IRS_CR_8,IRS_CR_9} consider fully-digital beamforming (FDB) at the CBS in RIS-aided CR systems. Due to the large number of antennas in a mmWave MIMO system, these fully digital TPCs are inefficient in mmWave CR systems due to their requirement of a large number of RF chains, which are costly and power thirsty. Therefore, the recently proposed hybrid TPC \cite{HBF_3_1,HBF_8,HBF_3_2,HBF_4,HBF_6,HBF_7,HBF_13,HBF_14,HBF_18,HBF_19} has a higher efficiency in mmWave MIMO CR systems, where the TPC relies on a much lower number of RF chains. Specifically, Wang \textit{et al.} \cite{HBF_3_1} proposed hybrid beamforming (HBF) for a single-user mmWave MIMO system, where the low resolution phase shifters of the RF TPC and RC pair are designed successively for maximizing the SE of the system. Furthermore, the baseband (BB) TPC and RC are obtained based on the effective BB channel for further improving the SE. As a further advance, Zhan \textit{et al.} \cite{HBF_4} consider a MU, multi-stream mmWave MIMO system, for which they propose zero-forcing (ZF) and successive interference cancellation (SIC)-based HBF to deal with both the multi-user interference (MUI) and inter-stream interference (ISI).
The authors of \cite{HBF_8} proposed a discrete Fourier transform (DFT)-aided user clustering aided hybrid TPC by considering both partially- and fully-connected architectures in mmWave MIMO systems. They also quantified the energy efficiency (EE) for both the proposed architectures and concluded that the partially-connected architecture has a higher EE efficiency. 
Moreover, Zhang \textit{et al.} \cite{HBF_7} proposed an energy-efficient hybrid TCP and RC based on block diagonalization for the MU mmWave MIMO downlink. To improve the EE, the authors have proposed a water-filling solution for optimizing the power, which maximizes the weighted sum-SE of the system. 
As a further advance, Chen \textit{et al.} \cite{HBF_19} proposed low-complexity hybrid TPC schemes based on orthogonal frequency-division multiplexing (OFDM) for wideband mmWave multi-user (MU) MIMO systems.
In the context of HBF-aided mmWave MIMO CR systems, Tsinos \textit{et al.} \cite{HBF_13} proposed a hybrid TPC and RC design for mmWave MIMO CR systems while considering both TP, IP and unit modulus constraints for their hybrid architecture. This design is based on the alternating direction method of multipliers (ADMM) method considering full CSI at both the CBS and SUs. Moreover, our work in \cite{HBF_14} relied on limited CSI to design the hybrid TPC/RC of the mmWave MIMO CR downlink, which maximizes the sum-SE of the secondary system. However, the sum-SE metric results in the problem of low user fairness. Since the users having high channel-quality enjoy a high rate, while those having low-quality channels may have rates close to zero. 
To circumvent this problem, the paper also proposed hybrid TPC and RC designs for maximizing the geometric mean of the SU's rate in \cite{HBF_14}. Our recent work in \cite{HBF_15} investigates hybrid TPC/RC designs conceived for a frequency selective mmWave MIMO CR system, while considering practical uniform rectangular planar arrays (URPAs) both at the CBS and the SUs.  

The authors of \cite{IRS_mm_6,IRS_mm_8,IRS_mm_1,IRS_mm_3,IRS_mm_4,IRS_mm_9,IRS_mm_11,IRS_mm_10,IRS_mm_7,HBF_20} consider a joint active hybrid TPC design at the transmitter or base station (BS) and passive beamforming at the RIS in an RIS-aided mmWave MIMO system. 
More specifically, Bahingayi \textit{et al.} \cite{IRS_mm_6} consider a RIS-aided single-user mmWave MIMO system and formulate a problem to optimize the RIS RM. They employ singular value decomposition (SVD) of the channel and a heuristic greedy search method for determining the array response vectors that maximize the SE. Furthermore, they solved the RM optimization problem using the  Riemannian conjugate gradient (RCG) algorithm. Li \textit{et al.} \cite{IRS_mm_3} proposed a joint active hybrid TPC at the BS and a passive beamformer at the RIS for minimizing the total TP at the BS, while considering a quality of service (QoS) constraint for each user in the RIS-aided mmWave MU MIMO downlink. They used the RCG algorithm for handling the constant magnitude constraints on the elements of the RF TPC and RIS RM. Furthermore, Gong \textit{et al.} \cite{IRS_mm_4} proposed a joint active hybrid TPC and passive beamformer for an RIS-aided mmWave MU MIMO system to minimize the MSE. They conceived an accelerated RCG algorithm based on the majorization minimization (MM) method for addressing the non-convex unit modulus constraint on the elements of the RF TPC and RIS RM. Niu \textit{et al.} \cite{IRS_mm_9} have considered a double RIS-aided MU mmWave MIMO system and proposed joint hybrid TPC and passive beamforming design for maximizing the weighted sum-rate of the system under specific QoS constraints. They used the BCD method to design the BB TPC by employing quadratically constrained quadratic programming (QCQP), while the RIS RM was optimized using a price-mechanism-based RCG algorithm.
Pradhan \textit{et al.} \cite{IRS_mm_1} proposed joint active hybrid TPC designs for employment at the BS and passive beamformer designs at the RIS to minimize the mean squared error (MSE) in RIS-aided mmWave MU MIMO systems. They have leveraged a gradient projection method to deal with the non-convex unit modulus constraints imposed on the elements of the RF TPC and RIS RM.
Furthermore, Cheng \textit{et al.} \cite{IRS_mm_11} consider a beam-steering codebook to capture the practical implementation of a finite-resolution RF TPC and RM with limited feedback in the RIS-aided mmWave MU MIMO downlink. In their work, the authors have derived an upper bound for the achievable rate imposed by the finite resolution of the codebook and the limited feedback. 
As a further advance, Hong \textit{et al.} \cite{IRS_mm_7} exploited the sparsity of the angular domain in mmWave MIMO channels to jointly design the active hybrid TPC of the BS and the passive beamformer of the RIS for both narrowband and wideband RIS-aided mmWave MIMO systems. Moreover, Chen \textit{et al.} \cite{HBF_20} investigated the effect of beam squint in RIS-aided mmWave wideband systems. The authors therein proposed a novel technique for mitigating the beam squint effect via optimization of the passive RM.
However, none of the contributions reviewed above have conceived hybrid TPC and RC solutions for RIS-aided mmWave MIMO CR systems. This motivates us to consider the underlay RIS-assisted mmWave MIMO CR downlink. The novel contributions of this work are boldly contrasted to the existing studies in Table \ref{tab:lit_rev} at a glance. The detailed contributions of this paper are discussed next.  
\subsection{Contributions of this work}\label{contributions}
\begin{itemize}
\item Explicitly, this is the first paper to analyze the benefits of using an RIS in the mmWave MIMO CR downlink, where a CBS transmits multiple streams to multiple SUs in the presence of a PU. We formulate a sum-SE maximization problem for the given CR system to design the hybrid transceiver and passive RM under the TP, IP, and the non-convex unity modulus constraints on the elements of the RF TPC, RF RCs and RM. The problem formulated is highly non-convex and not tractable due to the non-convex constraints as well as owing to the coupling of variables in the objective function (OF) and constraints. To solve this problem, we transform the problem into a tractable one by employing a two-stage hybrid TPC design approach. Furthermore, we propose a BCD algorithm for the design of the RF TPC, RF RC and RM, alternatively.
\item For a fixed RM, we decompose the RF TPC and RF RC design problem into a series of sub-problems, where we formulate the optimization problem to design the pair of RF TPC and RF RC vectors. In order to optimize each sub-problem successively, we propose the successive Riemannian conjugate gradient algorithm, where each pair of RF TPC and RF RC vectors are optimized jointly, which is suitable for large-scale optimization. Similarly, for a fixed RF TPC and RF RC, each element of the RM is optimized successively based on the RCG algorithm. 
\item To design the BB TPC and BB RC we propose a pair of methods termed: D-SVD and P-SVD, which are based on the effective BB channel to maximize the sum-SE of the system. The D-SVD method directly uses the SVD of the direct effective BB channel of the SUs, whereas the P-SVD method uses the SVD of the channel projected onto the null-space of the PU's channel. Subsequently, this is followed by design of the BB TPC using the ZF method and proportional optimal power allocation. 
\item Finally, simulation results are provided for quantifying the efficiency of the proposed methods for an RIS-aided mmWave MIMO system.
\end{itemize}
\subsection{Notation}\label{notation}
Boldface capital letters, boldface small letters, and normal typeface letters represent matrices, vectors, and scalar quantities, respectively. To denote $(i,j)$th element of matrix $\mathbf{A}$, we use the notation $\mathbf{A}{(i,j)}$; the Hermitian and conjugate transpose of a matrix $\mathbf{A}$ are denoted by $\mathbf{A}^H$ and $\mathbf{A}^*$, respectively; $\left\vert\left\vert \mathbf{A} \right\vert\right\vert_F$ denotes the the Frobenius norm of $\mathbf{A}$, whereas $\left\vert\mathbf{A}\right\vert$ represents its determinant; $\text{Tr}(\mathbf{A})$ denotes its trace; ${\left\vert\left\vert\mathbf{a}\right\vert\right\vert}_{p}$ represents $p$-th norm of $\mathbf{a}$; ${\cal D}(\mathbf{a})$ denotes a diagonal matrix with vector $\mathbf{a}$ on its main diagonal; $\mathbf{A} \odot\mathbf{B}$ is the Hadamard product of $\mathbf{A}$ and $\mathbf{B}$; $\nabla f(\mathbf{a}_i)$ denotes the gradient vector of function $f(\mathbf{a})$ at the point $\mathbf{a}_i$; the real part of a quantity is denoted by $\Re\{\cdot\}$; ${\mathbf I}_M$ denotes an $M \times M$ identity matrix; the symmetric complex Gaussian distribution of mean $\mathbf{a}$ and covariance matrix $\mathbf{A}$  is represented as ${\cal CN}(\mathbf{a}, \mathbf{A})$.
\section{\uppercase{RIS-aided mmWave MU MIMO downlink CR System}}\label{mmWave MU MIMO CR System}
\subsection{System model}\label{system model}
\begin{table}[t]
   \centering
     \caption{ ABBREVIATIONS} \label{tab:ABB}
   \begin{adjustbox}{width=0.8\linewidth}
\begin{tabular}{l r}
    \hline
     Abbreviations & Explanations \\ [0.5ex]
 \hline
 mmWave & Millimeter wave \\
 MIMO & Multiple-input multiple-output \\
 RIS & Reconfigurable intelligent surface \\
 CBS & Cognitive base station \\
 TA & Transmit antenna \\
 RA & Receive antenna \\
 TPC & Transmit precoder \\
 RC & Receive combiner \\
 RF & Radio frequency \\
 BB & Base band \\
 RM & Reflection matrix \\
 PU & Primary user \\
 SU & Secondary user \\
 SE & Spectral efficiency \\
 IP & Interference power \\
 TP & Transmit power \\
 SVD & Singular value decomposition \\
 BCD & Block coordinate descent \\
 RCG & Riemannian conjugate gradient \\
 \hline
\end{tabular}
\end{adjustbox}
\end{table}
\begin{figure}[t]
\centering
\includegraphics [width=8cm]{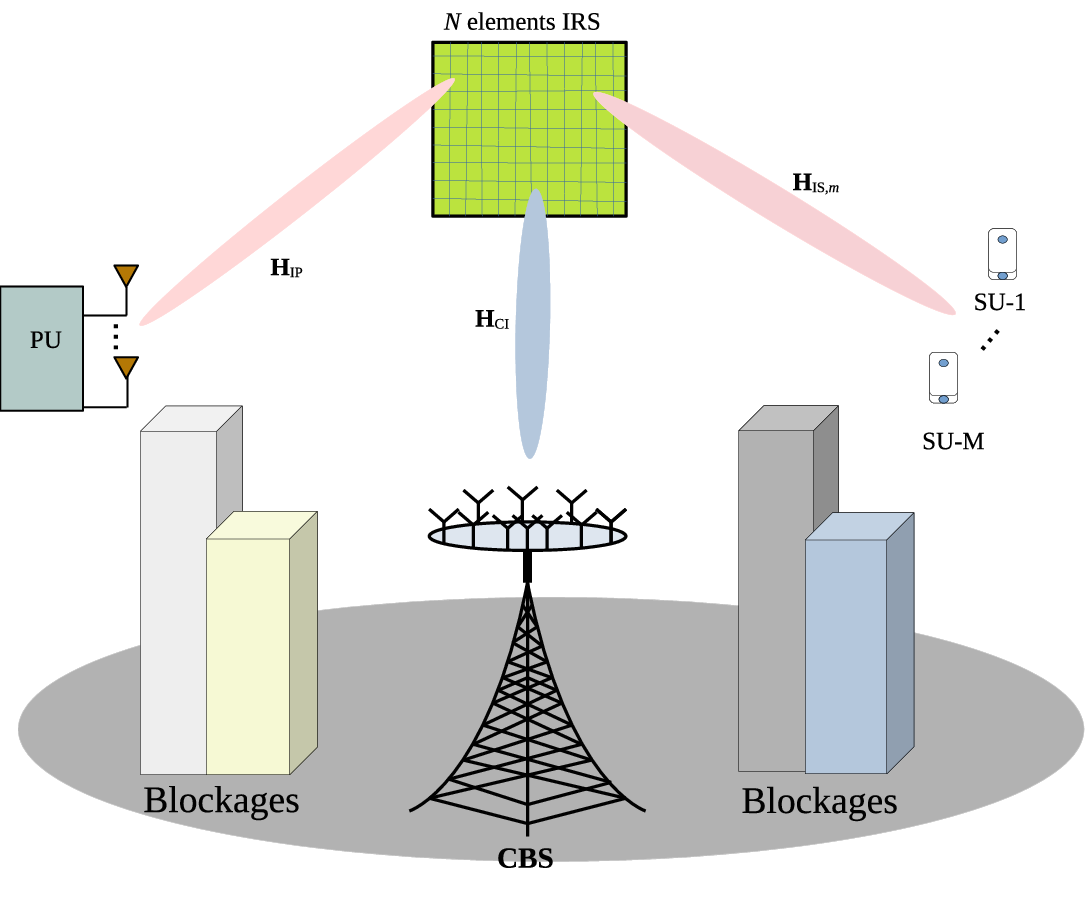}
\caption{RIS-aided mmWave MIMO downlink cognitive radio system.}
\label{fig:Fig1}
\vspace{-4mm}
\end{figure}
We consider the underlay RIS-aided mmWave MIMO CR downlink as shown in Fig. \ref{fig:Fig1}. A CBS equipped with $N_\mathrm{t}$ transmit antennas (TAs) and $M_\mathrm{t}$ RF chains is transmitting data to $M$ SUs in the presence of a PU with the aid of a RIS. 
Each SU and PU is equipped with $N_\mathrm{r}$ receive antennas (RAs) and $M_\mathrm{r}$ RF chains. A fully-connected hybrid TPC is employed at the CBS to transmit $MN_s$ data streams, i.e., $N_s$ data streams to each SU, where we have $MN_s\leq M_t$ and $N_s\leq M_r$. The signal vector $\mathbf{s}=\left[\mathbf{s}^T_1,\hdots,\mathbf{s}^T_M\right]^T \in \mathbb{C}^{MN_\mathrm{s}\times 1}$ is initially precoded by the BB TPC $\mathbf{F}_\mathrm{BB}=\left[\mathbf{F}_{\mathrm{BB},1},\hdots, \mathbf{F}_{\mathrm{BB},M}\right]\in \mathbb{C}^{M_\mathrm{t}\times MN_\mathrm{s}}$  followed by the RF TPC $\mathbf{F}_\mathrm{RF} \in \mathbb{C}^{M_\mathrm{t} \times N_\mathrm{t}}$, where $\mathbf{s}_m \in \mathbb{C}^{N_s \times 1}$ and $\mathbf{F}_{\mathrm{BB},m}\in \mathbb{C}^{M_\mathrm{t}\times N_\mathrm{s}}$ are the transmitted signal and the BB TPC corresponding to the $m$th SU. 
The RIS is deployed on the facade of a building close to the CBS and SUs for substantially suppressing interference at the PU due to the downlink transmission between the CBS and the SUs. In particular, the RIS comprises $N$ reflective elements and it is programmable and reconfigurable via an RIS controller.
Let us denote the RIS RM as $\boldsymbol{\Phi } = \mathcal{D}\left(\left[\phi_1, \hdots, \phi_n \right]\right)$ with $\phi_n = \alpha_n e^{j\theta_n}$, where $\alpha_n \in [0, 1]$ and $\theta_n \in [0, 2\pi]$ are the amplitude and phase shift of the $n\text{-th}$ reflective element. Assuming $\alpha_n=1$ for maximizing the reflection gain of the RIS leads to $\left\vert\boldsymbol{\Phi}(n,n)\right\vert=1$. The cascaded channel matrix corresponding to the $m$th SU is given by $\mathbf{H}_{m}=\mathbf{H}_{\mathrm{IS},m}\boldsymbol{\Phi}\mathbf {H}_{\mathrm{CI}} \in \mathbb{C}^{N_\mathrm{r} \times N_\mathrm{t}}$, where $\mathbf {H}_{\mathrm{CI}}\in \mathbb{C}^{N \times N_\mathrm{t}}$ and $\mathbf{H}_{\mathrm{IS},m}\in \mathbb{C}^{N_\mathrm{r} \times N}$ are the channel links spanning from the CBS to the RIS and from the RIS to the $m$th SU, respectively.  
Considering a frequency-flat block-fading channel, the signal $\mathbf{y}_m \in \mathbb{C}^{N_\mathrm{r} \times 1}$ received at the SU $m$ is given by
\begin{equation}\label{eqn:rx signal}
\begin{aligned}
\mathbf {y}_{m}=&\mathbf{ H}_m \mathbf{F}_{\rm RF}\mathbf{F}_{\rm BB}\mathcal{D}(\sqrt{\mathbf{p}})\mathbf{s} + \mathbf{n}_{m} \\
=&\mathbf{H}_m \mathbf{F}_{\rm RF}\mathbf{F}_{{\rm BB},m}\mathcal{D}(\sqrt{\mathbf{p}_m})\mathbf{s}_m \\
&+\sum_{n=1, n \neq m}^{M}\hspace{-0.3cm}\mathbf {H}_{m} \mathbf{F}_{\rm RF}\mathbf{F}_{{\rm BB},n} \mathcal{D}(\sqrt{\mathbf{p}_n})\mathbf{s}_{n}+  \mathbf{n}_{m}, 
\end{aligned}
\end{equation}
where $\mathbf{p} = \left[\mathbf{p}_1,\hdots,\mathbf{p}_M\right] \in \mathbb{C}^{MN_\mathrm{s} \times 1}$ represents the power allocation vector, and $\mathbf{p}_m(d)$ denotes the power assigned to the $d$th stream of SU $m$. Furthermore, $\mathbf{n}_m \in \mathbb{C}^{N_\mathrm{r}\times 1}$ is an additive white complex Gaussian noise process with distribution $\mathcal{CN}(\mathbf{0}, \sigma^2\mathbf{I})$. Upon considering a hybrid RC at each SU, the processed received signal $\widetilde{\mathbf{y}}_{m}\in \mathbb{C}^{N_\mathrm{s} \times 1}$ at the $m$th user is given by 
\begin{equation}
\begin{aligned}
\widetilde{\mathbf{y}}_{m}=  
  & \mathbf{W}_{{\rm BB},m}^H\mathbf{W}_{{\rm RF},m}^H\mathbf{H}_{m} \mathbf{F}_{\rm RF}\mathbf{F}_{{\rm BB},m}\mathcal{D}(\sqrt{\mathbf{p}_m})\mathbf{s}_{m} \nonumber
 \end{aligned}
\end{equation}
\begin{equation}\label{eqn:prc rx signal}
\begin{aligned}
  &+\sum_{n=1, n \neq m}^{M}\mathbf{W}_{{\rm BB},m}^H  
  \mathbf{W}_{{\rm RF},m}^H\mathbf{H}_{m} \mathbf{F}_{\rm RF}\mathbf{F}_{{\rm BB},n}\mathcal{D}(\sqrt{\mathbf{p}_n})\mathbf{s}_{n}\\ &+\mathbf{W}_{{\rm BB},m}^H\mathbf{W}_{{\rm RF},m}^H \mathbf{n}_{m},
 \end{aligned}
\end{equation}
where $\mathbf{W}_{{\rm RF},m} \in \mathbb{C}^{N_{\rm r} \times M_{\rm r}}$ and $\mathbf{W}_{{\rm BB},m}\in \mathbb{C}^{M_{\rm r} \times N_{\rm s}}$ are the RF RC and the BB RC matrices, respectively, of the $m$th SU. Moreover, the fully-connected hybrid antenna array at both the CBS and each SU ends leads to $\left\vert\mathbf{F}_{\rm RF}(i,j)\right\vert=1$ and $\left\vert\mathbf{W}_{{\rm RF},m}(i,j)\right\vert=1$. Similarly, the interference received at the PU is given by
\begin{equation}\label{eqn:rx signal_1}
\begin{aligned}
\mathbf{y}_P=&\mathbf{G} \mathbf{F}_{\rm RF}\mathbf{F}_{\rm BB}\mathcal{D}(\sqrt{\mathbf{p}})\mathbf{s}+\mathbf{n}_p,
\end{aligned}
\end{equation}
where $\mathbf{G}=\mathbf{H}_\mathrm{IP}\boldsymbol{\Phi }\mathbf{H}_\mathrm{CI}\in \mathbb{C}^{N_\mathrm{r} \times N_\mathrm{t}}$ is the effective channel matrix of the PU and $\mathbf{H}_\mathrm{IP}\in \mathbb{C}^{N_\mathrm{r} \times N}$ is the channel spanning from the CBS to the PU. 
\subsection{mmWave MIMO channel}\label{mmWave MIMO channel}
Throughout this paper, we employ the widely used Saleh-Valenzuela channel model \cite{IRS_mm_4,IRS_mm_9} of the wireless channel, which includes complex path losses, delays, angle-of-arrivals (AoAs), and angle-of-departures (AoDs). The frequency flat mmWave MIMO channel between two nodes is given by
\begin{equation}\label{eqn:channel}
\mathbf{H}_{i}= \sum_{l=1}^{N^{\rm p}_i}\alpha_{i,l}\mathbf{a}_{\rm r}(\phi^{\rm r}_{i,l},\theta^{\rm r}_{i,l})\mathbf{a}_{\rm t}^H(\phi^{\rm t}_{i,l},\theta^{\rm t}_{i,l}), 
\end{equation}
where the subscript $i\in \{\{{\rm CI}\}, \{{\rm IS},m\}, {\rm IP}\}$ represents the corresponding link, and $N^p_i$ denotes the number of multipath components in $\mathbf{H}_i$. The quantity $\alpha_{i,l}$ is the gain of the $l$th multipath component in $\mathbf{H}_i$. Furthermore, $\mathbf{a}_{\rm t}(\phi_{i,l}^t\theta_{i,l}^t)\in \mathbb{C}^{{{\rm col}(\mathbf{H}_i)}\times 1}$ denotes the transmit array response vector corresponding to the azimuth and elevation angles of departure (AoDs), namely $\phi_{i,l}^t$, $\theta_{i,l}^t$, respectively. Similarly, $\mathbf{a}_{\rm r}(\phi_{i,l}^r\theta_{i,l}^r)\in \mathbb{C}^{{{\rm row}(\mathbf{H}_i)}\times 1}$ denotes the receive array response vector corresponding to the azimuth and elevation angles of arrival (AoAs), namely $\phi_{i,l}^r$, $\theta_{i,l}^r$, respectively.
\begin{table*}[ht]
\begin{equation}\label{eqn:Gamma_m}
\mathbf{\Gamma}_m = \frac{\mathbf{W}_{{\rm BB},m}^H\mathbf{W}_{{\rm RF},m}^H\mathbf{ H}_{m} \mathbf{F}_{\rm RF}\mathbf{F}_{{\rm BB},m}{\cal D}(\mathbf{p}_m)\mathbf{F}_{{\rm BB},m}^H\mathbf{F}_{{\rm RF},m}^H\mathbf{ H}_{m}^H {\mathbf W}_{\rm RF}{\mathbf W}_{{\rm BB},m}}
{\sum_{n=1, n \neq m}^{M}{{\mathbf W}_{{\rm BB},m}^H{\mathbf W}_{{\rm RF},m}^H\mathbf{ H}_{m} \mathbf{F}_{\rm RF}\mathbf{F}_{{\rm BB},n}{\cal D}(\mathbf{p}_n)\mathbf{F}_{{\rm BB},n}^H\mathbf{F}_{{\rm RF}}^H\mathbf{H}_{m}^H \mathbf{W}_{{\rm RF},m}\mathbf{W}_{{\rm BB},m} + \sigma^2\mathbf{W}_{{\rm BB},m}^H\mathbf{W}_{{\rm RF},m}^H\mathbf{W}_{{\rm RF},m}\mathbf{W}_{{\rm BB},m}}}
\tag{8}
\end{equation}
\end{table*}
We consider uniform planar arrays (UPAs) at the BS, RIS, and at each UE. As a result, the array response vectors can be written as
\begin{equation}\label{eqn:array respo}
\begin{aligned} 
\mathbf{a}_z\left (\phi, \theta \right)=&\frac {1}{\sqrt{N_z}}\Biggl [1, \ldots, e^{j \frac {2 \pi }{\lambda } d \left ({o \sin \phi \sin \theta +p \cos \theta }\right)},\ldots, \\
&e^{j \frac {2 \pi }{\lambda } d \left ({(N_z^h-1) \sin \phi \sin \theta)+(N_z^v-1) \cos \theta }\right)}\Biggr]^{T},
\end{aligned}
\end{equation}
where $z\in \{\rm r,t\}$, $d$ is the antenna spacing or RIS element spacing, which is assumed to be half of the wavelength $\lambda$, $0\leq o< N_z^h$ and $0\leq p< N_z^v$, where $N_z^h$ and $N_z^v$ denote the number of horizontal and vertical elements of the UPA in the 2D plane, respectively.
\subsection{Problem formulation}\label{Problem Formulation}
This paper seeks to jointly design the hybrid TPC/RCs $\bigl\{\mathbf{ W}_{{\rm RF},m},\mathbf{ W}_{{\rm BB},m}\bigr\}_{m=1}^M$, $\mathbf{F}_{\rm RF},\mathbf{F}_{\rm BB}$, RIS RM $\mathbf{\Phi}$ and the power allocation vector $\mathbf{p}$ that maximizes sum-SE of the secondary system under TP, IP and the non-convex constant magnitude phase constraints. The corresponding SE of the SU $m$ is given by
\begin{equation}\label{eqn:R_sum}
{\cal R}_m = \log_{2}\Big(\big\vert \mathbf{I}_{N_{\rm s}}+\mathbf{\Gamma}_m \big\vert \Big),
\end{equation}
where the matrix $\mathbf{\Gamma}_m \in \mathbb{C}^{N_{\rm s} \times {N_{\rm s}}}$ represents the signal to interference plus noise ratio (SINR) power, which is given by Eq. (\ref{eqn:Gamma_m}) at the top of the next page. Moreover, due to downlink communication between the CBS and SUs in the same frequency band as the PU, the aggregate interference induced at the PU can be written as
\begin{equation}\label{eqn:I_PU}
\begin{aligned}
I_{PU} = \sum_{m=1}^{M}\big\vert\big\vert {\bf G}\mathbf{F}_{\rm RF}\mathbf{F}_{{\rm BB},m}{\cal D}(\sqrt{\mathbf{p}_m})\big\vert\big\vert^2_{F}.
\end{aligned}
\end{equation}
Therefore, for a given RIS-aided mmWave MIMO channel, the sum-SE of the downlink CR system can be formulated as
\begin{align}
&\mathcal{P}_{1}: \max_{\bigl\{\mathbf{ W}_{{\rm RF},m},\mathbf{ W}_{{\rm BB},m}\bigr\}_{m=1}^M,\mathbf{F}_{\rm RF},\mathbf{F}_{\rm BB},\mathbf{\Phi}, \mathbf{p}}\sum_{m=1}^{M}\mathcal{R}_m \label{eqn:system optimization} \tag{9a}\\ 
& \text {s.t.} \quad \left\vert\mathbf{F}_{\rm RF}(i,j)\right\vert = 1, \forall i, j, \label{eqn:system optimization_1_1}\tag{9b}\\
&\qquad\left\vert\mathbf{ W}_{{\rm RF},m}(i,j)\right\vert = 1, \forall i, j, m, \label{eqn:system optimization_1_2}\tag{9c}\\
&\qquad\left\vert\mathbf{\Phi}(n,n)\right\vert =1, \forall n, \label{eqn:system optimization_1_3} \tag{9d}\\
&\qquad I_{PU}\leq I_{\rm th},\label{eqn:system optimization_1_4} \tag{9e}\\
&\qquad \|\mathbf{F}_\mathrm{RF}\mathbf{F}_\mathrm{BB}\mathcal{D}(\mathbf{\sqrt{p}})\|^2_F \leq P_\mathrm{T}, \label{eqn:system optimization_1_5} \tag{9f}
\end{align}
where $I_\mathrm{th}$ and $P_\mathrm{T}$ control the IP at PU and the TP at the CBS. It is important to highlight here that both the CBS and each SU require complete knowledge of the channel matrices $\mathbf{H}_m, \mathbf{H}_\mathrm{IP}$ and $\mathbf{H}_\mathrm{CI}$, which is a typical requirement in underlay CR systems \cite{HBF_13, HBF_14}. Moreover, the CSI required can be readily obtained via the transmission of training symbols followed by employing suitable channel estimation techniques, as discussed in \cite{HBF_21}.


Observe from $\mathcal{P}_1$ that the non-convex OF (\ref{eqn:system optimization}) and the non-convex unit modulus constraints (\ref{eqn:system optimization_1_1}), (\ref{eqn:system optimization_1_2}), (\ref{eqn:system optimization_1_3}) imposed on the RF TPC, RCs, and RM elements make the problem highly non-convex. Also observe that the TPC, RC matrices and RM are coupled in the OF (\ref{eqn:system optimization}) and IP constraint (\ref{eqn:system optimization_1_4}), which makes $\mathcal{P}_{1}$ even more challenging to solve. Therefore, in order to find the solution, we propose a two-stage hybrid transceiver design based on the BCD principle, which is discussed in the next section. 

\section{Two-stage hybrid transceiver design for RIS-aided mmWave MIMO CR downlink}\label{Solution}
In order to maximize the sum-SE of the system, we decompose the BB TPC $\mathbf{F}_\mathrm{BB}$ as $\mathbf{F}_\mathrm{BB}=\mathbf{F}^{1}_{\rm BB}\mathbf{F}^{2}_{\rm BB}$, where $\mathbf{F}^{1}_{\rm BB} = [\mathbf{F}^{1}_{{\rm BB},1}, \hdots,\mathbf{F}^{1}_{{\rm BB},m}, \hdots, \mathbf{F}^{1}_{{\rm BB},M}]\in \mathbb{C}^{M_{\rm t}\times MN_{\rm s}}$ and $\mathbf{F}^{2}_{\rm BB}= [\mathbf{F}^{2}_{{\rm BB},1}, \hdots,\mathbf{F}^{2}_{{\rm BB},m}\\, \hdots, \mathbf{F}^{2}_{{\rm BB},M}]\in \mathbb{C}^{N_{\rm s}\times MN_{\rm s}}$. The key idea behind this decomposition is to design $\mathbf{F}_\mathrm{RF}$ and $\mathbf{F}^1_\mathrm{BB}$ for jointly maximizing the sum-SE in the first stage while ignoring the MUI. Subsequently, $\mathbf{F}^2_{\mathrm{BB}}$ is designed in the second-stage for mitigating the MUI. Therefore, the updated sum-SE maximization problem can be recast as
\begin{equation}\label{eqn:system optimization_1}
\begin{aligned}
&\mathcal{P}_{2}: \max_{\bigl\{\mathbf{ W}_{{\rm RF},m},\mathbf{ W}_{{\rm BB},m}\bigr\}_{m=1}^M,\mathbf{F}_{\rm RF},\mathbf{F}^1_{\rm BB},\mathbf{F}^2_{\rm BB},\mathbf{\Phi}, \mathbf{p}}{\cal R}_{\rm sum} \\ 
& \text {s.t.} \quad \text{(\ref{eqn:system optimization_1_1}), (\ref{eqn:system optimization_1_2}), (\ref{eqn:system optimization_1_3}), (\ref{eqn:system optimization_1_4}), (\ref{eqn:system optimization_1_5})}.
\end{aligned}
\tag{10}
\end{equation}

To solve $\mathcal{P}_{2}$, we focus first on the joint design of the RF TPC $\mathbf{F}_\mathrm{RF}$, RCs $\{\mathbf{W}_{\mathrm{RF},m}\}_{m=1}^M$ and RM $\boldsymbol{\Phi}$ based on the BCD method. Next, $\mathbf{F}^1_{\rm BB}$ and $\{\mathbf{W}_{{\rm BB},m}\}_{m=1}^M$ are determined by maximizing the sum-SE based on the associated effective BB channel. Finally, we compute $\mathbf{F}^2_{\rm BB}$ followed by the optimal power allocation vector $\mathbf{p}$.
\subsection{Joint RF TPC, RC, and RIS RM design}
Upon assuming that the MUI can be eliminated in the second-stage of the TPC, one can approximate the rate of the $m$th SU $\mathcal{R}_m$ at high SNR as  
\begin{equation}\label{eqn:mutual_1}
\begin{aligned} 
\mathcal{R}_m & \approx \mathrm{log}_2 \Bigg (\bigg | \mathbf{R}_m^{-1} \mathbf {W}^H_{\mathrm{BB},m}\mathbf{W}^H_{\mathrm{RF},m} \mathbf{H}_m \mathbf{F}_\mathrm{RF}\mathbf{F}^1_{\mathrm{BB},m} \mathcal{D}(\mathbf{p}_m) \\ 
&\qquad \times(\mathbf{F}^1_{\mathrm{BB},m})^H \mathbf{F}_\mathrm{RF}^H \mathbf{H}_m^H \mathbf{W}_{\mathrm{RF},m}\mathbf{W}_{\mathrm{BB},m} \bigg | \Bigg), 
\end{aligned}
\tag{11}
\end{equation}
where $\mathbf{R}_m=\sigma^2\mathbf{W}_{{\rm BB},m}^H\mathbf{W}_{{\rm RF},m}^H\mathbf{W}_{{\rm RF},m}\mathbf{W}_{{\rm BB},m}$ is the effective noise at the $m$th SU.
Note that for a large number of TAs and RAs, the optimal RF TPC and RCs are approximately orthogonal, i.e., $\mathbf{F}^H_\mathrm{RF}\mathbf{F}_\mathrm{RF} \propto \mathbf{I}_{M_\mathrm{t}}$ and $\mathbf{W}^H_{\mathrm{RF},m} \mathbf{W}_{\mathrm{RF},m} \propto \mathbf{I}_{M_\mathrm{r}}, \forall m$. Furthermore, the hybrid TPC and RCs approach the optimal fully-digital TPC and RCs, respectively, which obey the approximation $\left(\mathbf{F}^1_\mathrm{BB}\right)^H \mathbf{F}^H_\mathrm{RF}\mathbf{F}_\mathrm{RF}\mathbf{F}^1_\mathrm{BB}  \approx I_{MN_\mathrm{s}}$ and $\left(\mathbf{W}_{\mathrm{BB},m}\right)^H \mathbf{W}^H_{\mathrm{RF},m}\mathbf{W}_{\mathrm{RF},m}\mathbf{W}_{\mathrm{BB},m}  \approx I_\mathrm{N_s}$. Following these facts, one can assume that the matrices $\mathbf{F}^1_{\mathrm{BB},m}$ and $\mathbf{W}_{\mathrm{BB},m}$ are orthogonal, i.e., $\left(\mathbf{F}^1_{\mathrm{BB},m}\right)^H \mathbf{F}^1_{\mathrm{BB},m} \propto  I_{MN_\mathrm{s}}$ and $\left(\mathbf{W}_{\mathrm{BB},m}\right)^H \mathbf{W}_{\mathrm{BB},m} \propto I_\mathrm{N_s}$. Therefore, (\ref{eqn:mutual_1}) can be approximated as
\begin{equation}
\begin{aligned} 
\mathcal{R}_m &\approx \mathrm{log}_2 \Bigg (\bigg | \frac{1}{\sigma^2}\mathbf{W}^H_{\mathrm{RF},m} \mathbf{H}_m \mathbf{F}_\mathrm{RF}\underbrace{\mathbf{F}^1_{\mathrm{BB},m}\mathcal{D}(\mathbf{p}_m) (\mathbf{F}^1_{\mathrm{BB},m})^H }_\text{$\approx \mathcal{D}({\mathbf{q}_m})$} \\ 
& \qquad \times \mathbf{F}_\mathrm{RF}^H \mathbf{H}_m^H \mathbf{W}_{\mathrm{RF},m} \bigg | \Bigg)\\
& =  \mathrm{log}_2 \Bigg (\bigg | \frac{1}{\sigma^2} \mathbf{W}^H_{\mathrm{RF},m} \mathbf{H}_m \mathbf{F}_\mathrm{RF} \mathcal{D}({\mathbf{q}_m}) \mathbf{F}_\mathrm{RF}^H \mathbf{H}_m^H \mathbf{W}_{\mathrm{RF},m} \bigg | \Bigg)\\
& =  \mathrm{log}_2 \Bigg (\bigg | \frac{1}{\sigma^2} \mathbf{F}_\mathrm{RF}^H \mathbf{H}_m^H \mathbf{W}_{\mathrm{RF},m}\mathbf{W}^H_{\mathrm{RF},m} \mathbf{H}_m \mathbf{F}_\mathrm{RF} \mathcal{D}({\mathbf{q}_m}) \bigg | \Bigg)\nonumber
\end{aligned}
\end{equation}
\begin{equation}\label{eqn:mutual_2}
\begin{aligned} 
& =  \mathrm{log}_2 \left(\left\vert\frac{\mathcal{D}({\mathbf{q}_m})}{\sigma ^2}\right\vert\right) \\
&\qquad\qquad+\mathrm{log}_2 \Bigg (\bigg | \mathbf{F}_\mathrm{RF}^H \mathbf{H}_m^H \mathbf{W}_{\mathrm{RF},m}\mathbf{W}^H_{\mathrm{RF},m} \mathbf{H}_m \mathbf{F}_\mathrm{RF} \bigg | \Bigg).
\end{aligned}
\tag{12}
\end{equation}
Let us define the SVD of $\mathbf{H}_m = \mathbf{U}_m \mathbf{\Sigma}_m \mathbf{V}_m^H$ as
\begin{equation}\label{eqn:SVD} 
\mathbf{H}_{m} =
\begin{bmatrix}
\widehat{\mathbf{U}}_{m}& \widecheck{\mathbf{U}}_{m}
\end{bmatrix}
\begin{bmatrix}
  \widehat{\mathbf{\Sigma}}_{m} & 0 \\ 
  0 & \widecheck{\mathbf{\Sigma}}_{m}
\end{bmatrix}
\begin{bmatrix}
\widehat{\mathbf{V}}_{m}& \widecheck{\mathbf{V}}_{m}
\end{bmatrix}^H,
\tag{13}
\end{equation}
where $\widehat{\mathbf{U}}_{m}$ contains the first $M_\mathrm{r}$ columns of $\mathbf{U}_m$,  $\widehat{\mathbf{\Sigma}}_{m}$ is comprised of the first $M_\mathrm{r}$ singular values of $\mathbf{\Sigma}_m$ and $\widehat{\mathbf{V}}_{m}$ contains the first $M_\mathrm{r}$ columns of $\mathbf{V}_m$. Upon assuming the effective rank of $\mathbf{H}_m$ to be equal the maximum number of data streams per SU, i.e., $M_\mathrm{r}$, one can approximate $\mathbf{H}_m$ as $\mathbf{H}_m \approx \widehat{\mathbf {H}}_m=\widehat{\mathbf{U}}_m \widehat{\mathbf{\Sigma}}_m \widehat{\mathbf{V}}_m^H$. Furthermore, one can write $\mathbf{F}_\mathrm{RF}$ as $\mathbf{F}_\mathrm{RF}=\left[\mathbf{F}_{\mathrm{RF},1}, \hdots, \mathbf{F}_{\mathrm{RF},m}, \hdots, \mathbf{F}_{\mathrm{RF},M_\mathrm{r}}\right]$, where $\mathbf{F}_{\mathrm{RF},m}\in \mathbb{C}^{N_\mathrm{t}\times M_\mathrm{r}}$. Upon exploiting 
the above facts, one can rewrite (\ref{eqn:mutual_2}) as
\begin{equation}
\begin{aligned}
\mathcal{R}_m &\approx \mathrm{log}_2 \left(\left\vert\frac{\mathcal{D}({\mathbf{q}_m})}{\sigma ^2}\right\vert\right)\\
&\qquad+\mathrm{log}_2 \Bigg (\bigg | \mathbf{F}_{\mathrm{RF},m}^H \widehat{\mathbf {H}}_m^H \mathbf{W}_{\mathrm{RF},m}\mathbf{W}^H_{\mathrm{RF},m} \widehat{\mathbf {H}}_m \mathbf{F}_{\mathrm{RF},m} \bigg | \Bigg)\nonumber
\end{aligned}
\end{equation}
\begin{equation}\label{eqn:mutual_2_1}
\begin{aligned}
&\overset{(b)}{\approx} \mathrm{log}_2 \left(\left\vert\frac{\mathcal{D}({\mathbf{q}_m})}{\sigma ^2}\right\vert\right)+2\times \mathrm{log}_2 \Bigg (\bigg |\mathbf{W}^H_{\mathrm{RF},m} \widehat{\mathbf {H}}_m \mathbf {F}_{\mathrm{RF},m}\bigg | \Bigg), 
\end{aligned}
\tag{14}
\end{equation}
where $(a)$ follows since $\vert\mathbf{X}\mathbf{Y}\vert = \vert\mathbf{X}\vert\vert\mathbf{Y}\vert$ when $\mathbf{X}$ and $\mathbf{Y}$ are square matrices. As a result, the joint design of the RF TPC and RCs, and RM is formulated as:
\begin{equation}\label{eqn:mutual_3}
\begin{aligned} 
&\mathcal{P}_{3}:\\
&\max_{\hspace{-0.7mm}\left\{\mathbf{W}_{{\rm RF},m},\mathbf{F}_{{\rm RF},m}\right\}_{m=1}^M, \mathbf{\Phi}}\hspace{-0.5mm} \sum_{m = 1}^{M} \mathrm{log}_2 \hspace{-0.5mm}\bigg (\Big |\mathbf {W}^H_{\mathrm{RF},m} \mathbf{H}_{\mathrm{IS},m}\boldsymbol{\Phi}\mathbf {H}_{\mathrm{CI}} \mathbf{F}_{\mathrm{RF},m}\Big | \bigg) \\ 
 & \text {s.t.} \quad \text{(\ref{eqn:system optimization_1_1}), (\ref{eqn:system optimization_1_2}), (\ref{eqn:system optimization_1_3})}.
\end{aligned}
\tag{15}
\end{equation}
Note that the problem $\mathcal{P}_3$ is still intractable due to the non-convex unit modulus constraints. Additionally, the RF TPC $\mathbf{F}_{{\rm RF},m}$, RF RC $\mathbf{W}_{{\rm RF},m}$ and RM $\boldsymbol{\Phi}$ are coupled in the OF.
Hence, in order to solve this challenging problem, we propose the BCD-successive RCG (SRCG) algorithm, where $\mathbf{F}_{{\rm RF},m}, \mathbf{W}_{{\rm RF},m}$, and $\mathbf{\Phi}$ are designed alternatively by employing the RCG algorithm. As per this procedure, we initially jointly design the RF TPC $\mathbf{F}_{\mathrm{RF},m}$ and RF RCs $\mathbf{W}_{\mathrm{RF},m}$ for a fixed RM $\mathbf{\Phi}$. Subsequently, we design the RM $\mathbf{\Phi}$ for the $\mathbf{F}_\mathrm{RF}$ and $\mathbf{W}_{\mathrm{RF},m}$ computed in the previous step. The proposed BCD-SRCG algorithm in described next in detail.
\subsubsection{Optimization of RF TPC and RC}
For a fixed RM $\mathbf{\Phi}$, we seek to design $\mathbf{W}_{{\rm RF},m},\mathbf{F}_{{\rm RF},m}$ based on the known effective channel $\widehat{\mathbf{H}}_m$. As a result, the RF TPC and RF RC design problem can be formulated as
\begin{equation}\label{eqn:mutual_4}
\begin{aligned} 
\mathcal{P}_{4}:&\max_{\left\{\mathbf{W}_{{\rm RF},m},\mathbf{F}_{{\rm RF},m}\right\}_{m=1}^M} \sum_{m = 1}^{M}\mathrm{log}_2 \bigg (\Big |\mathbf {W}^H_{\mathrm{RF},m} \mathbf{H}_m \mathbf{F}_{\mathrm{RF},m}\Big | \bigg) \\ 
&\text {s.t.} \quad \text{(\ref{eqn:system optimization_1_1}), (\ref{eqn:system optimization_1_2})}.
\end{aligned}
\tag{16}
\end{equation}
To further solve the above problem, we develop a novel SRCG algorithm, where we decompose $\mathcal{P}_{4}$ into a series of sub-problems. Explicitly, each pair of the RF TPC and RF RC are designed successively by invoking the RCG algorithm.
Let us consider $\mathbf{F}_{\mathrm{RF},m}\overset{\Delta}{=}\left[\mathbf{f}_{\mathrm{RF},m,1}, \hdots, \mathbf{f}_{\mathrm{RF},m,M_\mathrm{r}}\right]$ and $\mathbf{W}_{\mathrm{RF},m}\overset{\Delta}{=}\left[\mathbf{w}_{\mathrm{RF},m,1}, \hdots, \mathbf{w}_{\mathrm{RF},m,M_\mathrm{r}}\right]$, where $\mathbf{f}_{\mathrm{RF},m,l}$ and $\mathbf{w}_{\mathrm{RF},m,l}$ are the $l$th columns of $\mathbf{F}_{\mathrm{RF},m}$ $\mathbf{W}_{\mathrm{RF},m}$, respectively. Let us define $\mathbf{F}_{\mathrm{RF},m,{\backslash l}}$ and $\mathbf{W}_{\mathrm{RF},m,{\backslash l}}$ as the matrices that exclude the vectors $\mathbf{f}_{\mathrm{RF},m,l}$ and $\mathbf{w}_{\mathrm{RF},m,l}$ from $\mathbf{F}_{\mathrm{RF},m}$ and $\mathbf{W}_{\mathrm{RF},m}$, respectively. 
\begin{figure*}[t]
\begin{align} 
&\sum_{m = 1}^{M}\mathrm{log}_2 \bigg (\Big |\mathbf {W}^H_{\mathrm{RF},m} \mathbf{H}_m \mathbf{F}_{\mathrm{RF},m}\Big | \bigg)=\sum_{m = 1}^{M}\mathrm{log}_2 \bigg (\Big |\mathbf{W}^H_{\mathrm{RF},m} \widehat{\mathbf{U}}_m\widehat{\mathbf{\Sigma}}_m\widehat{\mathbf{V}}_m^H \mathbf {F}_{\mathrm{RF},m}\Big | \bigg)=\sum_{m = 1}^{M}\mathrm{log}_2 \bigg (\Big |\widehat{\mathbf{\Sigma}}_m\widehat{\mathbf{V}}_m^H \mathbf{F}_{\mathrm{RF},m}\mathbf{W}^H_{\mathrm{RF},m}\widehat{\mathbf{U}}_m\Big | \bigg)\label{successive_1}\tag{17}\\ 
&=\sum_{m = 1}^{M}\mathrm{log}_2 \Bigg (\bigg |\widehat{\mathbf{\Sigma}}_m\widehat{\mathbf{V}}_m^H \left[\mathbf{F}_{\mathrm{RF},m,{\backslash l}}\,\mathbf {f}_{\mathrm{RF},m,l}\right] \left[\mathbf {W}_{\mathrm{RF},m,{\backslash l}}\, \mathbf{w}_{\mathrm{RF},m,l}\right]^H \widehat{\mathbf {U}}_m\bigg | \Bigg)\tag{18}\\ 
&=\sum_{m = 1}^{M}\mathrm{log}_2 \Bigg (\bigg |\widehat{\mathbf{\Sigma}}_m\widehat{\mathbf{V}}_m^H \mathbf{F}_{\mathrm{RF},m,{\backslash l}}\mathbf{W}_{\mathrm{RF},m,{\backslash l}}^H\widehat{\mathbf {U}}_m + \widehat{\mathbf{\Sigma }}_m\widehat{\mathbf{V}}_m^H\mathbf{f}_{\mathrm{RF},m,l} \mathbf {w}_{\mathrm{RF},m,l}^H \widehat{\mathbf{U}}_m \bigg | \Bigg)\tag{19}\\
&\approx \sum_{m = 1}^{M}\mathrm{log}_2 \Bigg (\bigg |\left(\widehat{\mathbf{\Sigma }}_m\widehat{\mathbf{V}}_m^H \mathbf{F}_{\mathrm{RF},m,{\backslash l}}\mathbf{W}_{\mathrm{RF},m,{\backslash l}}^H\widehat{\mathbf{U}}_m\right) \Big [ \mathbf {I}_{M_\mathrm{r}}+\left(\alpha \mathbf {I}_{M_\mathrm{r}}+\widehat{\mathbf{\Sigma }}_m\widehat{\mathbf{V}}_m^H \mathbf{F}_{\mathrm{RF},m,{\backslash l}}\mathbf{W}_{\mathrm{RF},m,{\backslash l}}^H\widehat{\mathbf{U}}_m\right)^{-1} \nonumber\\ 
&\qquad\qquad\qquad\qquad\qquad\qquad\qquad\qquad\qquad\qquad\qquad\qquad\qquad\qquad\qquad\qquad\qquad \times \widehat{\mathbf{\Sigma}}_m\widehat{\mathbf {V}}_m^H\mathbf{f}_{\mathrm{RF},m,l}\mathbf{w}_{\mathrm{RF},m,l}^H \widehat{\mathbf{U}}_m\Big ] \bigg | \Bigg) \tag{20}\\ 
&=\sum_{m = 1}^{M}\mathrm{log}_2 \Bigg (\bigg | \widehat{\mathbf{\Sigma}}_m\widehat{\mathbf{V}}_m^H \mathbf{F}_{\mathrm{RF},m,{\backslash l}}\mathbf{W}_{\mathrm{RF},m,{\backslash l}}^H\widehat{\mathbf {U}}_m \bigg | \Bigg) + \sum_{m = 1}^{M}\mathrm{log}_2 \Bigg (\bigg |\Big [ \mathbf {I}_{M_\mathrm{r}}+\left(\alpha \mathbf {I}_{M_\mathrm{r}}+\widehat{\mathbf{\Sigma}}_m\widehat{\mathbf {V}}_m^H \mathbf{F}_{\mathrm{RF},m,{\backslash l}}\mathbf{W}_{\mathrm{RF},m,{\backslash l}}^H \widehat{\mathbf{U}}_m\right)^{-1} \nonumber \\ 
&\qquad\qquad\qquad\qquad\qquad\qquad\qquad\qquad\qquad\qquad\qquad\qquad\qquad\qquad\qquad\qquad\qquad\times \widehat{\mathbf{\Sigma }}_m\widehat{\mathbf{V}}_m^H\mathbf {f}_{\mathrm{RF},m,l}\mathbf{w}_{\mathrm{RF},m,l}^H \widehat{\mathbf{U}}_m\Big ] \bigg | \Bigg)\tag{21}\\
&=\sum_{m = 1}^{M}\mathrm{log}_2 \Bigg (\bigg | \mathbf {W}_{\mathrm{RF},m,{\backslash l}}^H \widehat{\mathbf {U}}_m \widehat{\mathbf{\Sigma}}_m\widehat{\mathbf{V}}_m^H \mathbf{F}_{\mathrm{RF},m,{\backslash l}} \bigg | \Bigg) + \sum_{m = 1}^{M}\mathrm{log}_2 \Bigg (\bigg |\Big [ 1+\mathbf{w}_{\mathrm{RF},m,l}^H\widehat{\mathbf {U}}_m\left(\alpha \mathbf{I}_{M_\mathrm{r}}+\widehat{\mathbf{\Sigma }}_m\widehat{\mathbf{V}}_m^H \mathbf{F}_{\mathrm{RF},m,{\backslash l}} \right. \nonumber\\
& \left.\quad\qquad\qquad\qquad\qquad\qquad\qquad\qquad\qquad\qquad\qquad\qquad\qquad\qquad\qquad\qquad\qquad\times\mathbf{W}_{\mathrm{RF},m,{\backslash l}}^H\widehat{\mathbf {U}}_m\right)^{-1}\widehat{\mathbf{\Sigma}}_m\widehat{\mathbf {V}}_m^H\mathbf{f}_{\mathrm{RF},m,l}\Big ] \bigg | \Bigg) \label{successive_2}\tag{22}
\end{align}
\normalsize
\hrulefill
\vspace*{4pt}
\end{figure*}
Following the steps from (\ref{successive_1}) to (\ref{successive_2}) shown at the top of the next page, the OF of (\ref{eqn:mutual_4}) can be reformulated as 
\begin{equation}\label{eqn:mutual_6}
\begin{aligned} 
&\sum_{m = 1}^{M}\mathrm{log}_2 \bigg (\Big |\mathbf {W}^H_{\mathrm{RF},m} \mathbf{H}_m \mathbf{F}_{\mathrm{RF},m}\Big | \bigg) \approx \sum_{m = 1}^{M}\mathrm{log}_2 \Bigg (\bigg | \mathbf{W}_{\mathrm{RF},m,{\backslash l}}^H \mathbf {H}_m \\
&\times \mathbf {F}_{\mathrm{RF},m,{\backslash l}} \bigg | \Bigg) 
+\sum_{m = 1}^{M}\mathrm{log}_2 \Bigg (\bigg\vert \left[1+ \mathbf{w}_{\mathrm{RF},m,l}^H \mathbf{Q}_{m,l}\mathbf{f}_{\mathrm{RF},m,l} \right] \bigg\vert\Bigg), 
\end{aligned}
\tag{23}
\end{equation}
where the matrix $\mathbf{Q}_{m,l} \in \mathbb{C}^{N_\mathrm{r}\times N_\mathrm{t}}$ is defined as
\begin{equation} \label{eqn:Q_l}
\begin{aligned}
\mathbf {Q}_{m,l} \triangleq &\widehat{\mathbf {U}}_m\left(\alpha \mathbf {I}_{M_\mathrm{r}}+\widehat{\mathbf{\Sigma}}_m\widehat{\mathbf{V}}_m^H \mathbf{F}_{\mathrm{RF},m,{\backslash l}} \mathbf{W}_{\mathrm{RF},m,{\backslash l}}^H\widehat{\mathbf{U}}_m\right)^{-1} \\
&\qquad\qquad\qquad\qquad\qquad\qquad\qquad\times \widehat{\mathbf{\Sigma}}_m\widehat{\mathbf{V}}_m^H . 
\end{aligned}
\tag{24}
\end{equation}
Observe that when $\mathbf{F}_{\mathrm{RF},m,{\backslash l}}$ and $ \mathbf{W}_{\mathrm{RF},m,{\backslash l}}$ are known, the first term of (\ref{eqn:mutual_6}) and $\mathbf {Q}_{m,l}$ are rendered constant. As a result, the sub-problem of optimizing of the RF TPC and RC reduces to the equivalent problem 
\begin{align} \label{eqn:mutual_7} 
&\mathcal{P}_{5}: \max_{\mathbf {w}_{\mathrm{RF},m,l},\mathbf {f}_{\mathrm{RF},m,l}} \sum_{m = 1}^{M}\mathrm{log}_2 \Bigg (\Big | \left[1+ \mathbf{w}_{\mathrm{RF},m,l}^H \mathbf{Q}_{m,l}\mathbf{f}_{\mathrm{RF},m,l} \right] \Big | \Bigg) \tag{25}\\ 
& \text {s.t.} \quad\left\vert\mathbf {f}_{\mathrm{RF},m,l}(i)\right\vert =1, \, i=1,\ldots, N_t, \label{eqn:mutual_7_a}\tag{25a}\\
& \qquad\left\vert\mathbf {w}_{\mathrm{RF},m,l}(j)\right\vert =1, \, j=1,\ldots, N_r. \label{eqn:mutual_7_b} \tag{25b}
\end{align} 
Furthermore, the OF in the above equation can be upper-bounded using Jensen's inequality as
\begin{equation} \label{eqn:mutual_8} 
\begin{aligned} 
& \sum_{m = 1}^{M}\mathrm{log}_2 \Bigg (\Big | \left[1+ \mathbf{w}_{\mathrm{RF},m,l}^H \mathbf{Q}_{m,l}\mathbf{f}_{\mathrm{RF},m,l} \right] \Big | \Bigg) \\
& \leq \mathrm{log}_2 \Bigg ( \left[1+ \sum_{m = 1}^{M} \Big |\mathbf{w}_{\mathrm{RF},m,l}^H \mathbf{Q}_{m,l}\mathbf{f}_{\mathrm{RF},m,l} \Big |\right]  \Bigg).
\end{aligned} 
\tag{26}
\end{equation}
We maximize the upper bound of (\ref{eqn:mutual_7}), and the corresponding optimization problem is given by
\begin{equation}\label{eqn:mutual_9} 
\begin{aligned} 
&\mathcal{P}_{6}:\max_{\mathbf {w}_{\mathrm{RF},m,l},\mathbf {f}_{\mathrm{RF},m,l}} \mathrm{log}_2 \Bigg ( \left[1+ \sum_{m = 1}^{M} \Big |\mathbf{w}_{\mathrm{RF},m,l}^H \mathbf{Q}_{m,l}\mathbf{f}_{\mathrm{RF},m,l} \Big |\right]  \Bigg) \\ 
& \text {s.t.} \quad \text{(\ref{eqn:mutual_7_a}), (\ref{eqn:mutual_7_b})}.
\end{aligned} 
\tag{27}
\end{equation}
Observe that the entries of $\mathbf {w}_{\mathrm{RF},m,l}$ and $\mathbf {f}_{\mathrm{RF},m,l}$ are subject to unit-modulus constraints. Thus, in order to jointly design each pair of $\mathbf {w}_{\mathrm{RF},m,l}$ and $\mathbf {f}_{\mathrm{RF},m,l}$, we concatenate them as $\mathbf{z}_{m,l} = \left[\mathbf {w}_{\mathrm{RF},m,l}^H, \mathbf{f}_{\mathrm{RF},m,l}^H\right]^H \in \mathbb{C}^{(N_\mathrm{r}+N_\mathrm{t})\times 1}$ to form a higher-dimensional vector, which is subject to the unit-modulus constraint. As a result, $\mathcal{P}_6$ can be reformulated as
\begin{equation}\label{eqn:RCG_1}
\begin{aligned}
\mathcal{P}_{7}:\qquad\qquad&\max_{\mathbf{z}_{m,l}}\sum_{m = 1}^{M} \left\vert\mathbf {z}_{m,l}^H \mathbf {D}_{m,l} \mathbf{z}_{m,l}\right \vert \\
&{\text {s.t. }} \quad |\mathbf{z}_{m,l}(i)| = 1,\quad \forall i, m, l,
\end{aligned}
\tag{28}
\end{equation}
where $\mathbf{D}_{m,l}=\begin{bmatrix} \mathbf{I}_{N_\mathrm{r}\times N_\mathrm{r}} \\ \mathbf{0}_{ N_\mathrm{t} \times N_\mathrm{r}}  \end{bmatrix} \mathbf{Q}_{m,l} \begin{bmatrix} \mathbf{0}_{N_\mathrm{t} \times N_\mathrm{r}} \mathbf{I}_{N_\mathrm{t} \times N_\mathrm{t}}  \end{bmatrix} \in \mathbb{C}^{(N_\mathrm{r}+N_\mathrm{t}) \times (N_\mathrm{r}+N_\mathrm{t})}$. 
Note that $\mathcal{P}_7$ is also non-convex due to the non-convex unit modulus constraint imposed on each element of $\mathbf{z}_{m,l}$. To this end, let us define the feasible set $\mathcal{Z}$ for (\ref{eqn:RCG_1}) on the complex circle manifold as
\begin{equation} 
\mathcal {Z} = \left\{\mathbf{z}_{m,l} \in {\mathbb {C}^{(N_\mathrm{r}+N_\mathrm{t})\times 1}}: \qquad\left \vert\mathbf{z}_{m,l}(i)\right\vert =  1,\quad \forall i \right \}.
\tag{29}
\end{equation}
Therefore, the problem (\ref{eqn:RCG_1}) can be recast as
\begin{equation}
\begin{aligned}
&\max_{\mathbf{z}_{m,l}}f(\mathbf {z}_{m,l})=\sum_{m = 1}^{M} \left\vert\mathbf {z}_{m,l}^H \mathbf {D}_{m,l} \mathbf{z}_{m,l}\right \vert \\
&{\text {s.t. }} \quad \mathbf{z}_m(i) \in \mathcal{Z},\quad \forall i, m.
\end{aligned}
\tag{30}
\end{equation}
Furthermore, the Euclidean gradient of $f(\mathbf {z}_{m,l})$ is given by
\begin{equation}\label{RCG_3}
\begin{aligned}
\nabla f(\mathbf {z}_{m,l})=2\times \left [{\begin{array}{c} \mathbf {Q}_{m,l} \mathbf {f}_{\mathrm{RF},m,l}\ \\ 
 \mathbf{Q}^H_{m,l}\mathbf{w}_{\mathrm{RF},m,l} \end{array}}\right].
\end{aligned}
\tag{31}
\end{equation}
The RCG algorithm takes advantage of the Riemannian gradient to evaluate the descent direction, which is defined as the orthogonal projection of $\nabla f(\mathbf {z}_{m,l})$ onto the tangent space $T_{\mathbf{z}^i_{m,l}}\mathcal{Z}$ of the manifold $\mathcal{Z}$ at the associated point $\mathbf{z}^i_{m,l}$. This is mathematically expressed as
\begin{equation}\label{RCG_2}
\begin{aligned}
&T_{\mathbf {z}^i_{m,l}} \mathcal {Z}\\
&=\left \{{\mathbf{z}_{m,l} \in \mathbb {C}^{(N_\mathrm{r}+N_\mathrm{t})}: \Re\left \{{\mathbf{z}_{m,l} \odot (\mathbf{z}^i_{m,l})^{*}}\right \}=\mathbf {0}_{(N_\mathrm{r}+N_\mathrm{t})}}\right \}.
\end{aligned}
\tag{32}
\end{equation}
Subsequently, the Riemannian gradient at the point $\mathbf{z}^i_{m,l}$ is obtained as
\begin{equation} \label{RCG_4}
\begin{aligned}
\mathrm {grad}~ f(\mathbf{z}^i_{m,l})=\nabla f(\mathbf{z}^i_{m,l}) -\Re\left \{{\nabla f(\mathbf{z}^i_{m,l}) \odot (\mathbf{z}^i_{m,l})^{*}}\right \} \odot \mathbf{z}^i_{m,l}.
\end{aligned}
\tag{33}
\end{equation}
Similar to the conjugate gradient method of the Euclidean space, the update rule of the search direction in the manifold space is given by 
\begin{equation} \label{RCG_5}
\begin{aligned}
\boldsymbol {\eta }^{i+1}=-\mathrm {grad}~ f(\mathbf {z}^{i+1}_{m,l})+\lambda_1 \mathcal {T}_{\mathbf{z}^i_{m,l} \rightarrow \mathbf{z}^{i+1}_{m,l}}\left ({\boldsymbol {\eta }^{i}}\right),
\end{aligned}
\tag{34}
\end{equation}
where $\boldsymbol {\eta}^{i}$ denotes the search direction at $\mathbf {z}^{i}_{m,l}$, $\lambda_1$ is the update parameter choosen as the the Polak-Ribiere parameter \cite{RCG_1}, and $\mathcal {T}_{\mathbf {z}^i_{m} \rightarrow \mathbf {z}^{i+1}_m}\left ({\boldsymbol {\eta}^{i}}\right)$ represents the transport operation. Briefly, the transport operation is required because both $\boldsymbol {\eta }^{i+1}$ and $\boldsymbol {\eta }^i$ are in different tangent spaces and operations such as the sum in (\ref{RCG_5}) cannot be carried out directly. Therefore, the transport operation $\mathcal {T}_{\mathbf {z}^i_{m,l} \rightarrow \mathbf {z}^{i+1}_{m,l}}\left ({\boldsymbol {\eta}^{i}}\right)$ proposed in \cite{IRS_mm_3} is required to map the tangent vector at the previous search direction to its original tangent space of the current point $\mathbf{z}^{i+1}_{m,l}$, which is given by
\begin{equation}\label{RCG_6}
\begin{aligned} 
\mathcal {T}_{\mathbf {z}^i_{m,l} \rightarrow \mathbf {z}^{i+1}_{m,l}}\left ({\boldsymbol {\eta }^{i}}\right): & T_{\mathbf {z}^{i}_{m,l}} \mathcal {Z} \mapsto T_{\mathbf {z}^{i+1}_{m,l}} \mathcal {Z}: \\
& \,\,\,\, \boldsymbol {\eta }^{i} \mapsto \boldsymbol {\eta }^{i}-\Re\left \{{\boldsymbol {\eta }^{i} \odot (\mathbf {z}^{i+1}_{m,l})^{*}}\right \} \odot \mathbf {z}^{i+1}_{m,l}.
\end{aligned}
\tag{35}
\end{equation}
Upon determining the search direction $\boldsymbol {\eta }^{i+1}$, the retraction operation $\mathrm {Retr}_{\mathbf {z}^i_m}(\lambda _{2} \boldsymbol {\eta }_i)$ of \cite{IRS_mm_3} is performed for determining the destination on the manifold. Specifically,  $\mathrm {Retr}_{\mathbf {z}^i_m}(\lambda _{2} \boldsymbol {\eta }_i)$ maps the point on the tangent space $T_{\mathbf {z}^i_m}\mathcal{Z}$ to the manifold $\mathcal{Z}$, which is given by
\begin{equation}\label{RCG_7}
\begin{aligned} 
\mathrm {Retr}_{\mathbf {z}^i_{m,l}}(\lambda_2 \boldsymbol {\eta}^i): T_{\mathbf {z}^{i}_{m,l}} \mathcal {Z}\mapsto&\mathcal {Z}: \\ 
\lambda_{2} \boldsymbol {\eta }^{i}\mapsto&\frac {\left ({\mathbf {z}^{i}_{m,l}+\lambda_{2} \boldsymbol {\eta}^{i}}\right)_{j}}{\left |{\left ({\mathbf {z}^{i}_{m,l}+\lambda_{2} \boldsymbol {\eta }^{i}}\right)_{j}}\right |},
\end{aligned}
\tag{36}
\end{equation}
where $\lambda_{2}$ is the Armijo backtracking line search step size \cite{RCG_1} and $\left ({\mathbf {z}^{i}_{m,l}+\lambda_{2} \boldsymbol {\eta}^{i}}\right)_{j}$ denotes the $j$th entry of $\left ({\mathbf {z}^{i}_{m,l}+\lambda_{2} \boldsymbol {\eta}^{i}}\right)$. The key steps of the SRCG algorithm discussed above to solve problem (\ref{eqn:mutual_4}) are summarized in Algorithm 1.
\begin{algorithm}[t]\label{alg:RCG_algo}
\caption{SRCG Algorithm for solving (\ref{eqn:mutual_4}) }
\begin{algorithmic}[1]
\Require $\mathbf{H}_m$ and desired accuracy $\epsilon$ 
\State \textbf{Initialize:} $\mathbf{W}_{\mathrm{RF},m}$ $\mathbf{F}_{\mathrm{RF},m}$
        \For{$l=1: M_{\rm r}$ }
            \State Obtain $\mathbf{W}_{\mathrm{RF},m,{\backslash l}}$ from $\mathbf{W}_{\mathrm{RF},m}$, $\mathbf {F}_{\mathrm{RF},m,{\backslash l}}$ from $\mathbf{F}_{\mathrm{RF},m}$ and construct $\mathbf{z}_{m,l} = \left[\mathbf {w}_{\mathrm{RF},m,l}^H, \mathbf{f}_{\mathrm{RF},m,l}^H\right]^H$
            \State Update $\mathbf {Q}_{m,l} \triangleq \widehat{\mathbf {U}}_m(\alpha \mathbf {I}_{M_\mathrm{r}}+\widehat{\mathbf{\Sigma}}_m\widehat{\mathbf{V}}_m^H \mathbf{F}_{\mathrm{RF},m,{\backslash l}} \mathbf{W}_{\mathrm{RF},m,{\backslash l}}^H\widehat{\mathbf{U}}_m)^{-1} \widehat{\mathbf{\Sigma}}_m\widehat{\mathbf{V}}_m^H $    
        \State Calculate $\boldsymbol{\eta}^0 = -\mathrm {grad}~ f(\mathbf {z}^0_{m,l})$ according to (\ref{RCG_4}) and set $i = 0$;
             \While{$\left \vert \left \vert\mathrm {grad}~ f(\mathbf {z}^i_{m,l}) \right\vert\right\vert_f \leq \epsilon$}
              \State Choose the Armijo backtracking line search step size $\lambda_2$
              \State Obtain the next point $\mathbf{z}^{i+1}_{m,l}$ using retraction in (\ref{RCG_7})
              \State Determine Riemannian gradient $\mathrm {grad}~ f(\mathbf {z}^{i+1}_m)$ according to (\ref{RCG_4})
              \State Obtain the transport $\mathcal {T}_{\mathbf {z}_{m}^i \rightarrow \mathbf {z}^{i+1}_m}\left(\boldsymbol{\eta}^i\right)$ according to (\ref{RCG_6})
              \State Choose the Polak-Ribiere parameter $\lambda_1$
              \State Calculate the conjugate direction $\boldsymbol{\eta}^{i+1}$ according to (\ref{RCG_5})
              \State $i\leftarrow i+1$
              \EndWhile
              \State \textbf{end while}
        \State \textbf{update:} $\mathbf{W}_{\mathrm{RF},m}$ and  $\mathbf{F}_{\mathrm{RF},m}$   
    \EndFor
        \State {\bf end for}
    \State \textbf{return:} $\mathbf{W}_{\mathrm{RF},m}$ and $\mathbf{F}_{\mathrm{RF},m}$
\end{algorithmic}
\end{algorithm}

\subsubsection{Optimization of the RM $\mathbf{\Phi}$}
We now focus our attention on the design of the RIS RM $\mathbf{\Phi}$ for a fixed RF TPC $\mathbf{F}_{\mathrm{RF},m}$ and RF RC $\mathbf{W}_{\mathrm{RF},m}$, which maximize the sum-SE of the SUs. The pertinent problem of designing $\mathbf{\Phi}$ is given by
\begin{equation}\label{phi_1}
\begin{aligned} 
\mathcal{P}_{8}: \qquad\quad&\max_{\mathbf{\Phi}} \sum_{m = 1}^{M} \mathrm{log}_2 \bigg (\Big |\mathbf {W}^H_{\mathrm{RF},m} \mathbf{H}_{\mathrm{IS},m}\boldsymbol{\Phi}\mathbf {H}_{\mathrm{CI}} \mathbf{F}_{\mathrm{RF},m}\Big | \bigg) \\ 
& \text {s.t.}  \quad \text{(\ref{eqn:system optimization_1_3})}.
\end{aligned}
\tag{37}
\end{equation}
To solve this problem for a fixed $\mathbf {W}_{\mathrm{RF},m}$ and $\mathbf {F}_{\mathrm{RF},m}$, we once again adopt the successive optimization principle, where the problem (\ref{phi_1}) is decomposed into a series of sub-problems. In each sub-problem, $\phi_n$ is optimized for fixed values of the other $(N-1)$ elements. Toward this, let us define 
\begin{align}\label{phi_2}
\mathbf{R}_m\overset{\Delta}{=}&\mathbf {W}^H_{\mathrm{RF},m} \mathbf{H}_{\mathrm{IS},m}=\left[\mathbf{r}_{m,1}, \hdots, \mathbf{r}_{m,N} \right], \tag{38}\\
\mathbf{T}_m\overset{\Delta}{=}&\mathbf {H}_{\mathrm{CI}} \mathbf{F}_{\mathrm{RF},m}=\left[\mathbf{t}_{m,1}, \hdots, \mathbf{t}_{m,N} \right]^H, \tag{39}
\end{align}
where $\mathbf{r}_{m,n} \in \mathbb{C}^{M_\mathrm{r} \times 1}$ is the $n$th column of $\mathbf{R}_m$ and $\mathbf{t}^H_{m,n} \in \mathbb{C}^{M_\mathrm{t} \times 1}$ is the $n$th row of $\mathbf{T}_m$. Thus, the effective BB channel can be written as
\begin{equation}\label{phi_3}
\mathbf {W}^H_{\mathrm{RF},m} \mathbf{H}_{\mathrm{IS},m}\boldsymbol{\Phi}\mathbf {H}_{\mathrm{CI}} \mathbf{F}_{\mathrm{RF},m}\overset{\Delta}{=} \mathbf{R}_m \mathbf{\Phi} \mathbf{T}_m = \sum\limits_{n = 1}^N {{\phi_n}} \mathbf{r}_{m,n}\mathbf{t}^H_{m,n}.
\tag{40}
\end{equation}
Therefore, the OF of $\mathcal{P}_8$ can be rewritten as
\begin{equation}\label{phi_4}
\begin{aligned}
& \sum_{m = 1}^{M} \mathrm{log}_2 \bigg (\Big |\mathbf {W}^H_{\mathrm{RF},m} \mathbf{H}_{\mathrm{IS},m}\boldsymbol{\Phi}\mathbf {H}_{\mathrm{CI}} \mathbf{F}_\mathrm{RF}\Big | \bigg) \\
&=\sum_{m = 1}^{M} \mathrm{log}_2 \bigg (\Big |\sum\limits_{n = 1}^N {{\phi_n}} \mathbf{r}_{m,n}\mathbf{t}^H_{m,n}\Big | \bigg).
\end{aligned}
\tag{41}
\end{equation}
Furthermore, following the steps from (\ref{phi_succ_1}) to (\ref{phi_succ_2}) as shown at the top of the next page, (\ref{phi_4}) can be approximated as
\begin{equation}\label{phi_5}
\begin{aligned} 
& \sum_{m = 1}^{M}\log _2\left(\left\vert\sum\limits_{n = 1}^N {{\phi_n}} \mathbf{r}_{m,n}\mathbf{t}^H_{m,n}\right\vert\right) \\
&\approx \sum_{m = 1}^{M}{\log _2}\Big(\Big| \mathbf{\Delta}_{m,n} \Big| \Big) + \sum_{m = 1}^{M}{\log _2}\Big(\Big| {1 + {\phi_n}\delta_{m,n}} \Big| \Big),
\end{aligned}
\tag{46}
\end{equation}
where $\mathbf{\Delta}_{m,n}=\sum\limits_{i = 1,i \ne n}^N \phi_i \mathbf{r}_{m,i}{\mathbf{t}}_{m,i}^H$ and $\delta_{m,n}=\mathbf{t}^H_{m,n}\left(\alpha \mathbf{I}_{M_\mathrm{r}} + \sum\limits_{i = 1,i \ne n}^N \phi_i \mathbf{r}_{m,i} \mathbf{t}_{m,i}^H \right)^{-1}\mathbf{r}_{m,n}$. Observe that $\mathbf{\Delta}_{m,n}$ is fixed when the other $N-1$ reflective elements, RF TPC and RC are fixed. Therefore, the designed sub-problem for the optimization of the $n$th reflective element is given by 
\begin{equation}\label{phi_6}
\begin{aligned}
\mathcal{P}_{9}: \qquad\quad &\mathop {\max}\limits_{\phi_n} \quad\sum_{m = 1}^{M}{\log _2}\Big(\Big| {1 + {\phi_n}\delta_{m,n}} \Big| \Big) \\ 
&{\text {s.t. }} \quad \left\vert{\phi_n}\right\vert = 1,\forall n. 
\end{aligned}
\tag{47}
\end{equation}
Upon defining $\mathbf{\Psi}_n = \mathcal{D}\left(\left[\delta_{1,n},\hdots, \delta_{M,n}\right]\right) \in \mathbb{C}^{M \times M}$, $\mathcal{P}_9$ can be recast as
\begin{equation}\label{phi_7}
\begin{aligned}
\mathcal{P}_{10}: \qquad\quad&\max \limits_{\mathbf {\phi_n}} \quad  f(\phi_n) = \log _2\Big( {\Big| \mathbf{I}_M + \phi_n\mathbf{\Psi}_n \Big|} \Big) \\ 
&{\text {s.t. }} \quad  \left\vert{\phi_n}\right\vert = 1,\forall n. 
\end{aligned}
\tag{48}
\end{equation}
The unit modulus constraint on $\phi_n$ renders the above problem non-convex. To solve this problem, we once again adopt the above-mentioned RCG algorithm for designing the RF TPC and RC. To this end, the Euclidean gradient of the function $f(\phi_n)$ is formulated as
\begin{equation}\label{phi_8}
\begin{aligned}
\nabla f(\phi_n)=\mathrm{Tr}\left(\left({\mathbf{I}_M + {\phi_n}\mathbf{\Psi}_n}\right)^{-1}\mathbf{\Psi}_n \right).
\end{aligned}
\tag{49}
\end{equation}
Therefore, the problem (\ref{phi_5}) can be efficiently solved again by the RCG algorithm.
\begin{figure*}[t]
\begin{align} 
\sum_{m = 1}^{M}{\log _2}&\left( {\left| \sum\limits_{n = 1}^N {{\phi_n}} \mathbf{r}_{m,n}\mathbf{t}^H_{m,n}. \right|} \right) = \sum_{m = 1}^{M}{\log _2}\left( {\left| {\sum\limits_{i = 1,i \ne n}^N \phi_i\mathbf{r}_{m,i}\mathbf{t}_{m,i}^H + {\phi_n}\mathbf{r}_{m,n}\mathbf{t}^H_{m,n}} \right|} \right)\label{phi_succ_1} \tag{42}\\ 
& \approx \sum_{m = 1}^{M}{\log _2}\left(\left\vert \left( {\sum\limits_{i = 1,i \ne n}^N \phi_i\mathbf{r}_{m,i}{\mathbf{t}}_{m,i}^H} \right) \left(\mathbf{I}_{M_\mathrm{r}} + \left(\alpha \mathbf{I}_{M_\mathrm{r}} + \sum\limits_{i = 1,i \ne n}^N \phi_i \mathbf{r}_{m,i} {\mathbf{t}}_{m,i}^H \right)^{-1}\phi_n\mathbf{r}_{m,n}\mathbf{t}_k^H \right) \right \vert\right) \tag{43}\\ 
& = \sum_{m = 1}^{M}{\log_2}\left( {\left| {\sum\limits_{i = 1,i \ne n}^N \phi_i\mathbf{r}_{m,i} \mathbf{t}_i^H} \right|} \right)+\sum_{m = 1}^{M}{\log _2}\Bigg( \Bigg| \mathbf{I}_{M_\mathrm{r}} + \left( \alpha \mathbf{I}_{M_\mathrm{r}} + \sum\limits_{i = 1,i \ne n}^N \phi_i \mathbf{r}_{m,i}{\mathbf{t}}_{m,i}^H \right)^{-1}  {\phi_n}\mathbf{r}_{m,n}\mathbf{t}^H_{m,n} \Bigg| \Bigg) \tag{44}\\ 
& = \sum_{m = 1}^{M}\log_2\left( \left| \sum\limits_{i = 1,i \ne n}^N \phi_i \mathbf{r}_{m,i}{\mathbf{t}}_{m,i}^H \right| \right)+ \sum_{m = 1}^{M}{\log _2}\left( {\left| {1 + {\phi_n}\mathbf{t}^H_{m,n}\left(\alpha \mathbf{I}_{M_\mathrm{r}} + \sum\limits_{i = 1,i \ne n}^N \phi_i \mathbf{r}_{m,i} \mathbf{t}_i^H \right)^{-1}\mathbf{r}_{m,n}} \right|} \right)\label{phi_succ_2} \tag{45}
\end{align}
\normalsize
\hrulefill
\vspace*{4pt}
\end{figure*}

In summary, the BCD-SRCG algorithm successively designs the $l$th beamformer pair $\mathbf{f}_{\mathrm{RF},m,l}$ and $\mathbf{w}_{\mathrm{RF},m,l}$ of the RF TPC and RF RC jointly for a fixed setting of the RM $\mathbf{\Phi}$ by solving (\ref{eqn:mutual_9}) employing the RCG algorithm. Subsequently, with the RF TPC and RF RC thus computed, each element of $\mathbf{\Phi}$ is successively optimized according to (\ref{phi_7}), based on the RCG algorithm. As per the BCD-SRCG algorithm described above, $\mathbf{F}_\mathrm{RF}$, $\{\mathbf{W}_\mathrm{RF}\}_{m=1}^M$ and $\mathbf{\Phi}$ are alternately designed until convergence is achieved. The key steps of the BCD-SRCG procedure are summarized in Algorithm \ref{alg:algo_2}.
\begin{algorithm}[t]
\caption{BCD-SRCG algorithm toward solving (\ref{eqn:mutual_3})}
\label{alg:algo_2}
\textbf{Input: $\mathbf {H}_\mathrm{CI}, \
\mathbf{H}_{\mathrm{IS},m}, m=1,\hdots, M $}
\begin{algorithmic}[1]
     \State \textbf{Initialize:} $\mathbf{\Phi}$
     \For{$m=1:M$ }
       \State Obtain $\mathbf{H}_{m}=\mathbf{H}_{\mathrm{IS},m}\boldsymbol{\Phi}\mathbf {H}_{\mathrm{CI}}$
             \State  Construct $\mathbf{W}_{\mathrm{RF},m}, \mathbf{F}_{\mathrm{RF},m}$ based on Algorithm 1  
        \EndFor
        \State {\bf end for}
        \State Obtain $\mathbf{R}_m=\mathbf{W}^H_{\mathrm{RF},m}\mathbf{H}_{\mathrm{IS},m}$ and
        $\mathbf{T}_m=\mathbf {H}_{\mathrm{CI}} \mathbf{F}_{\mathrm{RF},m}$ 
       \For{$n=1: N$ }
         \State Obtain $\delta_{m,n}=\mathbf{t}^H_{m,n}\left(\alpha \mathbf{I}_{M_\mathrm{r}} + \hspace{-0.4cm}\sum\limits_{i = 1,i \ne n}^N \phi_i \mathbf{r}_{m,i} \mathbf{t}_{m,i}^H \right)^{-1}\hspace{-0.2cm}\mathbf{r}_{m,n},$ $m=1,\hdots, M$
         \State Construct $\mathbf{\Psi}_n = \mathcal{D}\left(\left[\delta_{1,n},\hdots, \delta_{M,n}\right]\right)$
         \State Obtain $\phi_n$ by solving (\ref{phi_7})
       \EndFor
       \State {\bf end for}
    \State  Go to step 2, until convergence of $\mathbf{F}_{\mathrm{RF},m}, \mathbf{W}_{\mathrm{RF},m}$ and $\mathbf{\Phi}$ is achieved 
    \State \textbf{return:} $\{\mathbf{W}_{\mathrm{RF},m}, \mathbf{F}_{\mathrm{RF},m}\}_{m=1}^M$ and $\mathbf{\Phi}$ 
\end{algorithmic}
\end{algorithm}

\subsection{BB TPC, RCs and optimal power allocation}\label{BB TPC}
This subsection presents a procedure for the design of the BB TPC $\mathbf{F}_{\mathrm{BB},m}$, BB RC $\mathbf{W}_{\mathrm{BB},m}, \forall m$ and determines the optimal power allocation $\mathbf{p}_m$, $m=1,\hdots,M$, which maximizes the sum-SE and minimizes the MUI. This is achieved as per the optimization in $\mathcal{P}_1$, for fixed RF TPC, RCs and RM.
As seen from $\mathcal{P}_1$, $\mathbf{F}_{\mathrm{BB},m}$ and $\mathbf{p}_m$ are coupled in the total TP and IP constraints, given by (\ref{eqn:system optimization_1_4}) and (\ref{eqn:system optimization_1_5}), respectively. Therefore, it is difficult to solve this problem. To compute the solution to this challenging problem, we present a pair of approaches, viz., the direct-SVD (D-SVD) and projected-SVD (P-SVD) techniques, that focus respectively on the spatial multiplexing of the SUs and on the interference mitigation at the PU. Both these approaches are discussed next in detail.
\subsubsection{D-SVD}
In this approach, the BB TPC and RCs are designed for maximizing the sum-SE based on the SVD of the channels $\mathbf{H}_m,\forall m$, while the power allocation is done to meet the TP and IP constraints. Therefore, the optimal fully-digital TPC and RC of the $m$th SU for the  D-SVD method are $\widehat{\mathbf{V}}_m$ and $\widehat{\mathbf{U}}_m$, respectively.
Furthermore, in the first-stage, $\mathbf{F}_{\rm RF}$ and $\mathbf{F}^1_{{\rm BB},m}$ are jointly designed for maximizing the sum-SE of the SUs. Therefore, for a fixed RF TPC $\mathbf{F}_\mathrm{RF}$ and RF RC $\mathbf{W}_{\mathrm{RF},m}$,  the quantities $\mathbf{F}^{1}_{\mathrm{BB},m}$ and $\mathbf {W}_\mathrm{{BB},m}$, that approach the optimal solution, are given by
\begin{align}
\mathbf{F}^{1,\mathrm{D}}_{\mathrm{BB},m} =&\left(\mathbf{F}^H_{\rm RF}\mathbf{F}_{\rm RF}\right)^{-1}\mathbf{F}^H_{\rm RF}\mathbf{V}^\mathrm{D}_m, \tag{50}\\ 
\mathbf{W}^\mathrm{D}_\mathrm{{BB},m} = &\left(\mathbf{W}^H_{{\rm RF},m}\mathbf{W}_{{\rm RF},m}\right)^{-1}\mathbf{W}^H_{{\rm RF},m}\mathbf{U}^\mathrm{D}_m, \tag{51}
\end{align}
where we have $\mathbf{V}^\mathrm{D}_m=\widehat{\mathbf{V}}_m$ and $\mathbf{U}^\mathrm{D}_m=\widehat{\mathbf{U}}_m$.
Furthermore, to mitigate the MUI, we use the ZF technique to design the precoder $\mathbf{F}^2_\mathrm{BB}$. As per this scheme, the CBS obtains the effective channel matrix of the $m$th SU as $\mathbf{H}^\mathrm{eff,D}_m = (\mathbf{W}^\mathrm{D}_{{\rm BB},m})^H\mathbf{W}_{{\rm RF},m}^H\mathbf{H}_{m} \mathbf{F}_{\rm RF}\mathbf{F}^{1,\mathrm{D}}_{{\rm BB},m} \in \mathbb{C}^{N_{\rm s}\times N_{\rm s}}, \forall m$ and stacks them as $\mathbf{\overline{H}}^\mathrm{D} =   
 \left[(\mathbf{H}^\mathrm{eff,D}_m)^T \hdots (\mathbf{H}^\mathrm{eff,D}_1)^T \hdots (\mathbf{H}^\mathrm{eff,D}_M)^T\right]^T\in \mathbb{C}^{MN_{\rm s}\times N_{\rm s}}$. Subsequently, the BB TPC $\mathbf{F}_{{\rm BB},2}$ is formulated as
 \begin{equation}\label{eqn:ZF}
 \begin{aligned}
\mathbf{F}^{2,\mathrm{D}}_{\rm BB} = \left((\mathbf{\overline{H}}^\mathrm{D})^H\mathbf{\overline{H}}^\mathrm{D}\right)^{-1}(\mathbf{\overline{H}}^\mathrm{D})^H.
\end{aligned}
\tag{52}
\end{equation}
Finally, the normalized BB TPC corresponding to the $m$th SU is given by 
\begin{equation}
\begin{aligned}
\mathbf{F}^\mathrm{D}_{{\rm BB},m} = \frac{\mathbf{F}^{1,\mathrm{D}}_{{\rm BB}, m} \mathbf{F}^{2,\mathrm{D}}_{{\rm BB}, m}}{\left\vert\left\vert \mathbf{F}_\mathrm{RF}\mathbf{F}^{1,\mathrm{D}}_{{\rm BB}, m} \mathbf{F}^{2,\mathrm{D}}_{{\rm BB}, m}\right\vert\right\vert}_F.
\end{aligned}
\tag{53}
\end{equation}
Furthermore, the resource allocation in a typical mmWave MU scenario can potentially suffer from unfairness due to coverage issues, differences in distance between the CBS and various SUs, as well as the priority of the SUs. Therefore, in order to avoid this, we introduce a weighted sum-SE for the system. The optimal power allocation problem of weighted sum-SE maximization, based on the D-SVD method, can now be formulated as
\begin{equation}
\begin{aligned}\label{eqn:P_opt}
&\mathcal{P}_{11}: \qquad \max_{\left\{{\mathbf{p}^\mathrm{D}_m}\right\}_{m=1}^{M}}\sum_{m=1}^M w^\mathrm{D}_m\mathcal{R}^\mathrm{D}_m \\ 
& \qquad \qquad \text {s.t.} \quad\text{(\ref{eqn:system optimization_1_4}), (\ref{eqn:system optimization_1_5})}, 
\end{aligned}
\tag{54}
\end{equation}
where $w^\mathrm{D}_m, \mathbf{p}^\mathrm{D}_m$ and $\mathcal{R}^\mathrm{D}_m$ denote the weight, power allocation and rate of the $m$th SU using the D-SVD method. Following Appendix \ref{AP_C}, the rate of the $m$th SU $\mathcal{R}^\mathrm{D}_m$, based on the D-SVD approach, can be simplified to the following expression
\begin{equation}\label{D_5}
\begin{aligned} 
\mathcal{R}^\mathrm{D}_m & \approx \log_{2}\Big(\Big\vert{\bf I}_{N_{\rm s}} + \frac{1}{\sigma^2_n}(\mathbf{F}^\mathrm{2,D}_{{\rm BB},m})^H\widehat{\mathbf{\Sigma}}^2_m\mathbf{F}^\mathrm{2,D}_{{\rm BB},m}{\cal D}(\mathbf{p}^\mathrm{D}_m)\Big\vert\Big).
\end{aligned}
\tag{55}
\end{equation}
Furthermore, let us now define the matrix $\mathbf{\Upsilon}^\mathrm{D}_m \in \mathbb{C}^{N_{\rm s}\times N_{\rm s}}$ as
 \begin{equation}\label{eqn:Y_m}
 \begin{aligned}
 \mathbf{\Upsilon}^\mathrm{D}_m  =& (\mathbf{F}^\mathrm{2,D}_{{\rm BB},m})^H\widehat{\mathbf{\Sigma}}^2_m\mathbf{F}^\mathrm{2,D}_{{\rm BB},m}, \\
 &{}\\
   \overset{(b)}{=}& \begin{bmatrix}
    \upsilon^\mathrm{D}_{m,1}\lVert \mathbf{f}^\mathrm{2,D}_{{\rm BB},m,1} \rVert_2^2 & \hdots & \mathbf{0} \\
    & \ddots & \\
    \mathbf{0}& \hdots & \upsilon^\mathrm{D}_{m,N_{\rm s}} \lVert  \mathbf{f}^\mathrm{2,D}_{{\rm BB},m,N_\mathrm{s}} \rVert_2^2
\end{bmatrix},
\end{aligned}
\tag{56}
 \end{equation}
where $\upsilon^\mathrm{D}_{m,i}$  represents the square of the $i$th principal diagonal element of the matrix $\widehat{\mathbf{\Sigma}}_m$ and $\mathbf{f}^\mathrm{2,D}_{{\rm BB},m,i}$ denotes the $i$th column of $\mathbf{F}^\mathrm{2,D}_{{\rm BB},m}$. Furthermore, the approximation $(b)$ employed in (\ref{eqn:Y_m}) follows by noting that the columns of $\mathbf{F}^\mathrm{2,D}_{{\rm BB},m}$ are asymptotically orthogonal for large antenna arrays \cite{HBF_1}. Furthermore, for the designed $\mathbf{\Phi}$, the IP constraint at the PU, due to the transmission by the CBS, can be formulated as
\begin{align}\label{eqn:IT}
I_{PU}&\leq I_{\rm th},\nonumber \\
\sum_{m=1}^{M}\text{Tr}\left(\mathbf {G}\mathbf{F}_{\rm RF}\mathbf{F}^\mathrm{D}_{{\bf BB},m}{\cal D}(\mathbf{p}^\mathrm{D}_m)\left(\mathbf{F}^\mathrm{D}_{{\bf BB},m}\right)^H\mathbf{F}_{\rm RF}^H\mathbf {G}^H\right)& \leq I_{\rm th},\nonumber\\
\sum_{m=1}^{M}\text{Tr}\Big({\cal D}(\mathbf{p}^\mathrm{D}_m)\underbrace{\left(\mathbf{F}^\mathrm{D}_{{\bf BB},m}\right)^H\mathbf{F}_{\rm RF}^H\mathbf {G}^H\mathbf {G}\mathbf{F}_{\rm RF}\mathbf{F}^\mathrm{D}_{{\rm BB},m}}_\text{$\mathbf{Z}_m$}\Big)& \leq I_{\rm th},\nonumber \\
\sum_{m=1}^{M}\sum_{d=1}^{N_{\rm s}}p^\mathrm{D}_{m,d}\zeta_{m,d}&\leq I_{\rm th},  \tag{57}
\end{align}
where $p^\mathrm{D}_{m,d}$ and $\zeta_{m,d}$ are the $d$th diagonal elements of ${\cal D}(\mathbf{p}^\mathrm{D}_m)$ and ${\mathbf Z_m}$, respectively. Similarly, the total TP constraint
at the CBS can be rewritten as 
\begin{equation}
\begin{aligned}
\sum_{m=1}^{M}\sum_{d=1}^{N_{\rm s}}p^\mathrm{D}_{m,d}t^\mathrm{D}_{m,d}&\leq P_\mathrm{T}, 
\end{aligned}
\tag{58}
\end{equation}
where $t^\mathrm{D}_{m,d}$ is the $d$th diagonal element of the matrix $\mathbf{T}^\mathrm{D}_m = \left(\mathbf{F}^\mathrm{D}_{\mathrm{BB},m}\right)^H\mathbf{F}^H_\mathrm{RF}\mathbf{F}_\mathrm{RF}\mathbf{F}^\mathrm{D}_{\mathrm{BB},m}$. Therefore, the sum-SE maximization of the system based on the D-SVD method is given by 
\begin{align}\label{eqn:DSVD_1}
 &\mathcal{P}_{12}:\nonumber \\ 
 &\mathop{\max}\limits_{p^\mathrm{D}_{m,d}} \quad \sum_{m=1}^{M}\sum_{d=1}^{N_{\rm s}}  \widetilde{w}^\mathrm{D}_{m,d}\log_{2}\left( 1 + \frac{\upsilon^\mathrm{D}_{m,d}\lVert \mathbf{f}^\mathrm{2,D}_{{\rm BB},m,d}\rVert_2^2}{\sigma^2} p^\mathrm{D}_{m,d}\right) \tag{59}\\
 & \ {\rm s.t.} \sum_{m=1}^{M}\sum_{d=1}^{N_{\rm s}}p^\mathrm{D}_{m,d}\zeta_{m,d}\leq I_{\rm th}, \label{eqn:DSVD_1_1}\tag{59a}\\ 
 &  \qquad \sum_{m=1}^{M}\sum_{d=1}^{N_{\rm s}}p^\mathrm{D}_{m,d}t^\mathrm{D}_{m,d}\leq P_\mathrm{T}, \label{eqn:DSVD_1_2} \tag{59b}\\
 &\qquad  p^\mathrm{D}_{m,d} \geq 0, \label{eqn:DSVD_1_3}\tag{59c}
 \end{align}
where $\widetilde{w}^\mathrm{D}_{m,d}$ is the weight corresponding to the $d$th stream of the $m$th SU. The theorem below quantifies the optimal power $p_{m,d}$ allocated to the $m$th SU and its $d$th stream.
\begin{theorem}\label{theorem 1}
The SE of the system given in $\mathcal{P}_{12}$ is maximized by
\begin{equation}\label{eqn:water filling}
\begin{aligned}
&p^\mathrm{D}_{m,d}\\
&=\max\left\{0, {\frac{1}{\lambda \zeta_{m,d}+\tau^\mathrm{D} t^\mathrm{D}_{m,d}}}-\frac{\sigma^2}{\widetilde{w}^\mathrm{D}_{m,d}\upsilon^\mathrm{D}_{m,d}\lVert \mathbf{f}^\mathrm{2,D}_{{\rm BB},m,d}\rVert_2^2}\right\} \forall m, d.
 \end{aligned}
 \tag{60}
\end{equation}
\begin{proof}
Given in Appendix \ref{AP_B}.
\end{proof}
\end{theorem}
The quantities $\lambda$ and $\tau^\mathrm{D}$ are the Lagrange multipliers associated with $\zeta_{m,d}$ and 
$t^\mathrm{D}_{m,d}$, respectively.
\subsubsection{P-SVD}
The P-SVD approach completely avoids interference at the PU due to communication between the CBS and SUs, which can be accomplished by projecting the channels of the SUs into the null space of the PU channel $\mathbf{G}$. To this end, let us define the SVD of $\mathbf{G}$ as
\begin{equation}
\mathbf{G}=\mathbf{U}_\mathrm{g}\mathbf{\Sigma}_\mathrm{g}\mathbf{V}_\mathrm{g}^H.
\tag{61}
\end{equation}
Therefore, after taking the projection, the effective channel of the $m$th SU as per this procedure is given by 
\begin{equation}
\widetilde{\mathbf{H}}_m=\mathbf{H}_m \left(\mathbf{I}_{N_\mathrm{t}}- \mathbf{V}_\mathrm{g}\mathbf{V}_\mathrm{g}^H\right).
\tag{62}
\end{equation}
In order to maximize the sum-SE of the system, the optimal TPC and RC can be found using the SVD of $\widetilde{\mathbf{H}}_m$. Let us define the SVD of $\widetilde{\mathbf{H}}_m$ as
\begin{equation}
\widetilde{\mathbf{H}}_m=\widetilde{\mathbf{U}}_m\widetilde{\mathbf{\Sigma}}_m\widetilde{\mathbf{V}}_m^H. 
\tag{63}
\end{equation}
Upon considering the optimal fully-digital TPC and RC for the $m$th SU as $\mathbf{V}^\mathrm{P}_m$ and $\mathbf{U}^\mathrm{P}_m$, which comprise the first $M_\mathrm{r}$ columns of $\widetilde{\mathbf{V}}_m$ and $\widetilde{\mathbf{U}}_m$, respectively, the BB TPC $\mathbf{F}^{1,\mathrm{P}}_{\mathrm{BB},m}$ and BB RC $\mathbf {W}^{\mathrm{P}}_{\mathrm{BB},m}$ for the P-SVD method are given by
\begin{align}
\mathbf{F}^{1,\mathrm{P}}_{\mathrm{BB},m}=&\left(\mathbf{F}^H_{\rm RF}\mathbf{F}_{\rm RF}\right)^{-1}\mathbf{F}^H_{\rm RF}\mathbf{V}^\mathrm{P}_m, \tag{64}\\ 
\mathbf {W}^{\mathrm{P}}_\mathrm{{BB},m}=&\left(\mathbf{W}^H_{\rm RF}\mathbf{W}_{\rm RF}\right)^{-1}\mathbf{W}^H_{\rm RF}\mathbf{U}^\mathrm{P}_m. \tag{65}
\end{align}
Furthermore, upon employing the ZF technique, $\mathbf{F}^{2,\mathrm{P}}_{\rm BB}$ is given by
\begin{equation}\label{eqn:ZF_2}
 \begin{aligned}
\mathbf{F}^{2,\mathrm{P}}_{\rm BB} = \left((\mathbf{\overline{H}^{\mathrm{P}}})^H\mathbf{\overline{H}}^{\mathrm{P}}\right)^{-1}(\mathbf{\overline{H}}^{\mathrm{P}})^H,
\end{aligned}
\tag{66}
\end{equation}
where $\mathbf{\overline{H}}^{\mathrm{P}} =   
 \left[(\mathbf{H}_1^{\rm eff,P})^T \hdots (\mathbf{H}_m^{\rm eff,P})^T \hdots (\mathbf{H}_M^{\rm eff,P})^T\right]^T\in \mathbb{C}^{MN_{\rm s}\times N_{\rm s}}$ and 
$\mathbf{H}^\mathrm{eff, P}_m = (\mathbf{W}^\mathrm{P}_{{\rm BB},m})^H\mathbf{W}_{{\rm RF},m}^H\\ \times\mathbf{H}_{m} \mathbf{F}_{\rm RF}\mathbf{F}^{1,\mathrm{P}}_{{\rm BB},m} \in \mathbb{C}^{N_{\rm s}\times N_{\rm s}}$.
Finally, the normalized BB TPC of the P-SVD method corresponding to the $m$th SU is given by 
\begin{equation}
 \begin{aligned}
\mathbf{F}^\mathrm{P}_{{\rm BB},m} = \frac{\mathbf{F}^{1, \mathrm{P}}_{{\rm BB}, m} \mathbf{F}^{2,\mathrm{P}}_{{\rm BB}, m}}{\left\vert\left\vert \mathbf{F}_\mathrm{RF}\mathbf{F}^{1, \mathrm{P}}_{{\rm BB}, m} \mathbf{F}^{2, \mathrm {P}}_{{\rm BB}, m}\right\vert\right\vert}_F.
\end{aligned}
\tag{67}
\end{equation}
Therefore, the sum-SE maximization for the RIS-aided mmWave MIMO CR downlink based on the P-SVD method is given by 
\begin{align}\label{eqn:final optimization}
 &\mathcal{P}_{13}: \mathop{\max}\limits_{p^\mathrm{P}_{m,d}} \quad \sum_{m=1}^{M}\sum_{d=1}^{N_{\rm s}}  \widetilde{w}^\mathrm{P}_{m,d}\log_{2}\left( 1 + \frac{\upsilon^\mathrm{P}_{m,d}\lVert \mathbf{f}^{2,\mathrm{P}}_{{\rm BB},m,d}\rVert_2^2}{\sigma^2} p^\mathrm{P}_{m,d}\right) \tag{68}\\ 
 & \ {\rm s.t.}\ \sum_{m=1}^{M}\sum_{d=1}^{N_{\rm s}}p^\mathrm{P}_{m,d}t^\mathrm{P}_{m,d}\leq P_\mathrm{T}, \tag{68a}\\
 &\qquad\qquad\qquad\quad\quad p^\mathrm{P}_{m,d} \geq 0, \tag{68b}
 \end{align}
where $\widetilde{w}^\mathrm{P}_{m,d}$ is the weight corresponding to the $d$th stream of
the $m$th SU, $\upsilon^\mathrm{P}_{m,d}$ denotes the square of the $d$th element on the principal diagonal of $\mathbf{\Sigma}^\mathrm{P}_m$, $\mathbf{f}^{2,\mathrm{P}}_{{\rm BB},m,d}$ is the $d$th column of the matrix $\mathbf{F}^{2,\mathrm{P}}_{{\rm BB},m}$ and $t^\mathrm{P}_{m,d}$ is the $d$th diagonal element of the matrix $\mathbf{T}^\mathrm{P}_m = (\mathbf{F}^\mathrm{P}_{\mathrm{BB},m})^H\mathbf{F}^H_\mathrm{RF}\mathbf{F}_\mathrm{RF}\mathbf{F}^\mathrm{P}_{\mathrm{BB},m}$.
Similar to Theorem 1, the sum SE of the system given in $\mathcal{P}_{13}$ based on the P-SVD method is maximized by the power allocation
\begin{equation}\label{eqn:water filling}
\begin{aligned}
p^\mathrm{P}_{m,d}=\max\left\{0, {{1}\over \tau^\mathrm{P} t^\mathrm{P}_{m,d}}-{{\sigma^{2}}\over{\widetilde{w}^\mathrm{P}_{m,d}\upsilon^\mathrm{P}_{m,d}\lVert \mathbf{f}^{2,\mathrm{P}}_{{\rm BB},m,d}\rVert_2^2}}\right\} \forall m, d,
 \end{aligned}
 \tag{69}
\end{equation}
where $\tau^\mathrm{P}$ is the Lagrange multiplier associated with $t^\mathrm{P}_{m,d}$.

Note that the proposed two-stage hybrid transceiver design relying on the optimization of the hybrid transceiver and passive RM can also be
applied in a wideband scenario by considering MIMO-orthogonal frequency division multiplexing (OFDM) modulation \cite{HBF_19,HBF_20}. 
In a MIMO-OFDM system, the BB TPC precedes the inverse fast Fourier transform (IFFT) operation, which is followed by the RF TPC at the transmitter side. On the other hand, the RF RC is succeeded by the fast Fourier transform (FFT) followed by the BB RC at each SU. Consequently, the BCD-SRCG algorithm can be extended to wideband scenarios for optimizing the RF TPC, RC, and passive RM, which are shared by all the subcarriers. The D-SVD and P-SVD methods can also be extended to optimize the BB TPCs, BB RCs, and power allocation for each subcarrier by employing the SVD of the corresponding frequency-selective mmWave MIMO channel.
\section{\uppercase{Simulation Results}}\label{simulation results}
\begin{figure}[t]
\centering
\includegraphics [width=8cm]{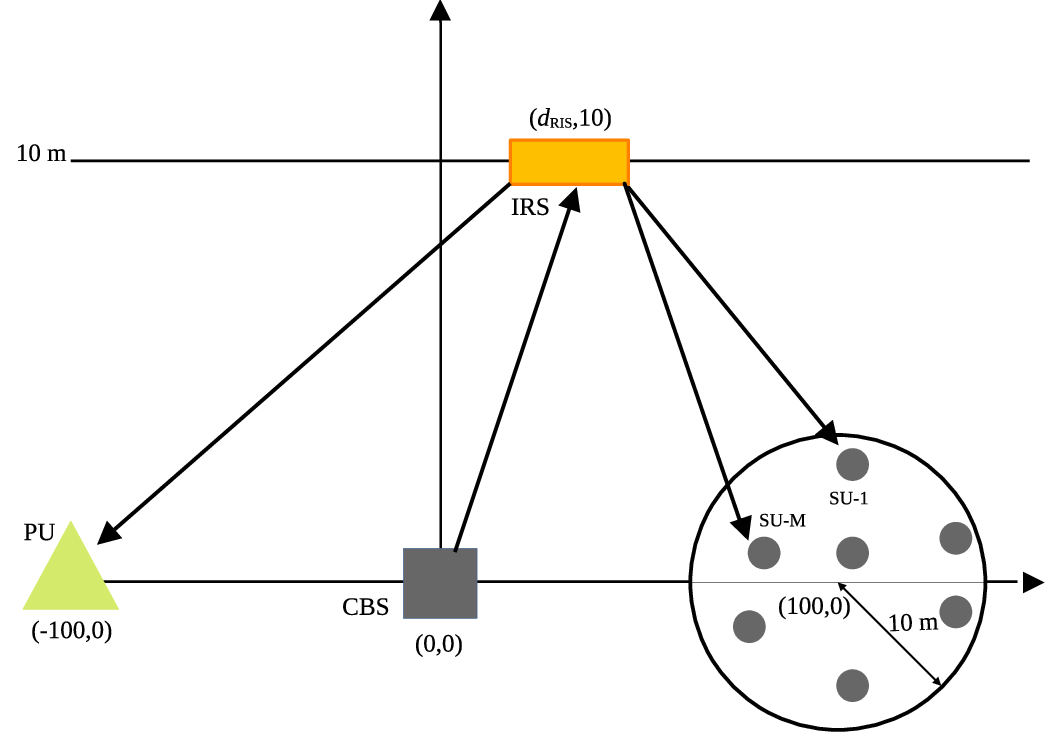}
\caption{Simulation setup for the RIS-aided mmWave MIMO CR downlink.}
\label{fig:Fig1_1}
\vspace{-4mm}
\end{figure}
In this section, we present simulation results for the algorithms proposed for jointly designing the hybrid transceiver and passive beamformer of an RIS-aided mmWave MIMO CR downlink. We consider a two-dimensional coordinated system to model the system as shown in Fig. \ref{fig:Fig1_1}, where $M$ SUs share the frequency band of a PU in a single cell. The CBS having a UPA structure is equipped with $N_\mathrm{t} = N_\mathrm{t_x}\times N_\mathrm{t_y}$ antennas and $M_\mathrm{t}=MM_\mathrm{r}$ RF chains and located at the origin $(0 \mathrm{m}, 0 \mathrm{m})$. Similarly, each SU and PU that have a UPA structure is equipped with $N_\mathrm{r} = N_\mathrm{r_x}\times N_\mathrm{r_y}$ antennas and $M_\mathrm{r}$ RF chains. The SUs are assumed to be uniformly distributed within a circle centered at $(100 \mathrm{m}, 0 \mathrm{m})$ and a radius of $10 $m, and the PU is situated at $(-100 \mathrm{m}, 0 \mathrm{m})$. Furthermore, the RIS is assumed to have $N$ reflective units with a UPA structure of $N = N_\mathrm{x}\times N_\mathrm{y}$ and situated at $(d_\mathrm{RIS} \mathrm{m}, 20 \mathrm{m})$. 
For the mmWave MIMO channel $\mathbf{H}_i$, the coefficients $\alpha_{i,l}$ are distributed independently, obeying the distribution as ${\cal CN}(0,\gamma_i^210^{-0.1PL(d_i)}), \forall l=\{1,\hdots, N^p_i\}$, where $\gamma_i=\sqrt{{\rm row}(\mathbf{H}_i){\rm col}(\mathbf{H}_i)/N^p_i}$ denotes the normalization factor. The quantity $PL(d_i)$ is the path-loss that depends on the distance $d_i$ associated with the corresponding link and it is modeled as \cite{IRS_mm_3}
\begin{equation}\label{eqn:path loss model}
\begin{aligned}
PL(d_i)\hspace{0.02in}[\rm dB] = \alpha + 10\beta\log_{10}(d_i)+\zeta,
\end{aligned}
\tag{70}
\end{equation}
where $\zeta \in {\cal CN}(0,\sigma_{\rm \zeta}^2)$. At the carrier frequency of $28$ GHz, the parameters of (\ref{eqn:path loss model}) are: $\alpha=61.4, \beta=2, \sigma_{\rm \zeta}=5.8 \hspace{0.02 in}{\rm dB}$ for LoS \cite{IRS_mm_3}. Moreover, we set the number of propagation paths to $N_i^{\rm p}=10, \forall i,$ with an angular spread of 10 degrees. The azimuth and elevation angles of departure and arrival follow a Laplacian distribution around the mean angle. The antenna spacing of both the CBS and of each SU is set to half-wavelength, i.e., $d_{\rm t}=d_{\rm r}=\frac{\lambda}{2}$. The noise variance $\sigma^2$ at each SU and PU is set to $-91\hspace{0.02 in}{\rm dBm}$. The simulation results are averaged over 500 independent channel realizations. The $\mathrm{SNR}$ is defined as ${\rm SNR}=\frac{P_{\rm t}}{\sigma^2}$, and its range is varied from $-10$ dB to $20$ dB to study the performance in both the low- and  high-$\mathrm{SNR}$ regions. The key simulation parameters are listed in Table \ref{Table2}. 
To demonstrate the efficiency of the proposed algorithms and to reveal some design insights, we compare the performance of the following algorithms when $N=16$ and $32$.
\begin{itemize}
    \item HBF (BCD-SRCG, D-SVD): This is the proposed BCD-SRCG algorithm and D-SVD approach for the joint hybrid transceiver and RIS RM design.
    \item HBF (BCD-SRCG, P-SVD): This is the proposed BCD-SRCG algorithm and P-SVD approach for the joint hybrid transceiver and RIS RM design.
    \item FDB w/o interference: For this scheme, CBS and each SU perform TPC and RC, respectively, using FDB, and the passive beamforming at RIS by employing the RCG approach, followed by power allocation without taking the IP constraint into account.
    \item HBF (Random Phase): The phases of the RM are assumed to be random and distributed uniformly between $0$ and $2\pi$, and the hybrid TPC/RC design is performed using the proposed SRCG and D-SVD algorithms.
    \item HBF (white spectrum): The joint hybrid TPC/RC and passive beamforming are performed using the BCD-SRCG algorithm, followed by equal power allocation to all streams of the SUs. 
\end{itemize}
We compare the performance by evaluating the achievable sum-SE of the SUs vs several important parameters, which are discussed next.
\begin{table}[t]
    \centering
    \caption{Key simulation parameters} \label{tab:simulation parameters}
\begin{tabular}{l r}\label{Table2}\\
\hline
Parameter & value \\ 
 \hline
 Carrier frequency & $28$ GHz\\
 Number of propagation paths, $N_i^{\mathrm p}$ & 10\\
 Path-loss parameters & $\alpha=61.4, \beta=2, \sigma_{\rm \zeta}=5.8$ dB \cite{IRS_mm_3}\\
 Noise power, $\sigma^2$ & $-91$ dBm\\
 Number of TAs, $N_\mathrm{t}$  & $128$\\
 Number of RAs, $N_\mathrm{r}$  & $8$\\
 Number of reflective elements, $N$ & $\{16, 32\}$ \\
 Location of RIS & $(20\mathrm{m}, 20\mathrm{m})$ \\
 Number of SUs, $M$ & $4$ \\
 Number of data streams per SU, $N_\mathrm{s}$ & $2$ \\
 Number of RF chains at each SU, $M_\mathrm{r}$ & $2$ \\
 Number of RF chains at CBS, $M_\mathrm{t}$ & $8$\\
 Interference threshold $(I_{\mathrm{th}})$ & $-10$dB to $20$dB \\
 $\mathrm{SNR}$, $\frac{P_\mathrm{t}}{\sigma^2}$ & $-10$dB to $20$dB \\
 \hline
\end{tabular}
\vspace{-4mm}
\end{table}
Unless otherwise stated, we consider an $8 \times 128$ system, where the CBS having $N_\mathrm{t} = 8 \times 16 = 128$  antennas and $M_\mathrm{t}=8$ RF chains is communicating with $M=4$ SUs, each having $N_\mathrm{r} = 2 \times 4 = 8$ antennas and $M_\mathrm{r}=2$ RF chains and $N=\{4\times 4=16, 4 \times 8=32\}$ reflective units, for a fixed IP threshold of $\Gamma=0$ dB.
\begin{figure}
\centering
\includegraphics[width = 9 cm]{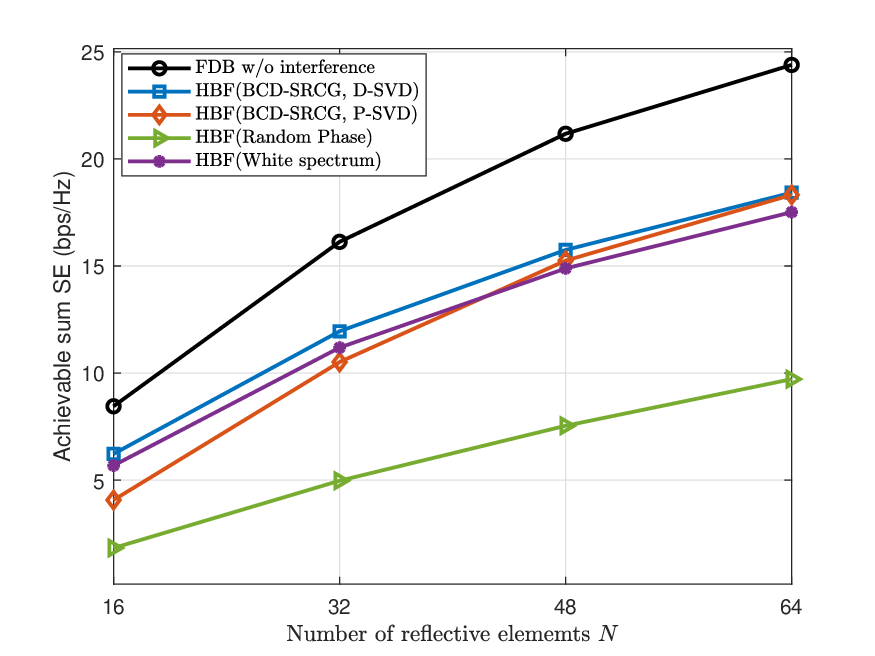}
\caption{Achievable sum-SE versus number of reflecting elements $N$ for an $8\times128$ RIS-aided mmWave MIMO CR system with SNR$=0$ dB and $\Gamma=0$ dB. All other parameters are listed in Table. \ref{tab:simulation parameters}.}
\label{fig:Fig2}
\vspace{-4mm}
\end{figure}
\begin{figure}
\centering
\includegraphics[width = 9 cm]{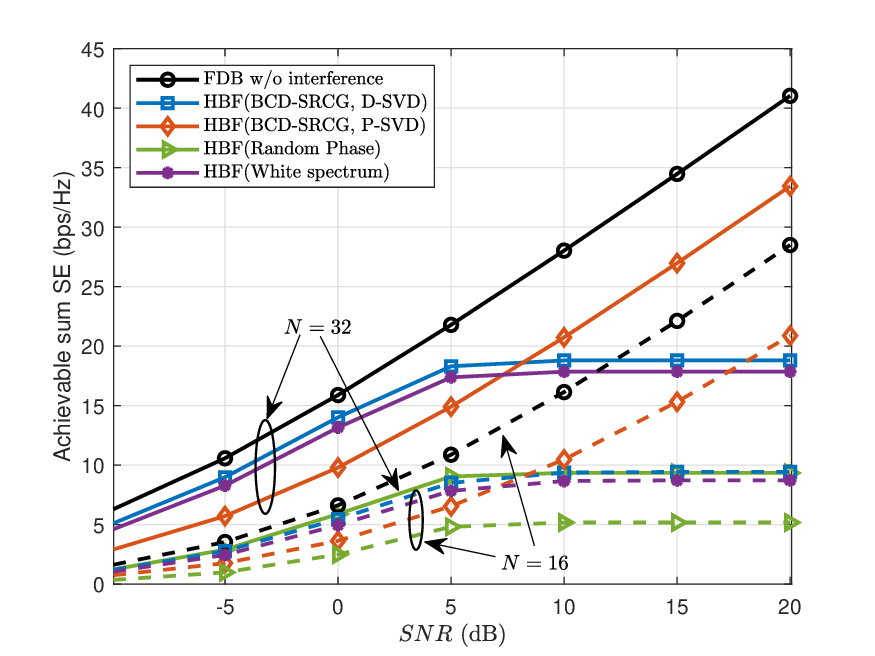}
\caption{Achievable sum-SE versus SNR for an $8\times128$ RIS-aided mmWave MIMO CR system with $\Gamma=0$ dB. All other parameters are listed in Table. \ref{tab:simulation parameters}.}
\label{fig:Fig3}
\vspace{-4mm}
\end{figure}
\subsubsection{Sum-SE versus $N$}
In Fig. \ref{fig:Fig2}, we plot the sum-SE vs. $N$ for fixed SNR$=0$ dB and $\Gamma=0$ dB. As seen from the figure, the sum-SE obtained using all the schemes increases with $N$ due to the higher passive beamforming gain. This demonstrates the advantages of introducing an RIS into mmWave MIMO CR systems. Moreover, the sum-SE of the proposed HBF (BCD-SRCG, D-SVD), HBF (BCD-SRCG, P-SVD) and HBF (white spectrum) schemes approach that of the FDB w/o interference and yield an improved performance in comparison to the HBF (Random Phase) approach. This demonstrates the effectiveness of our proposed joint hybrid TPC/RC and passive beamforming designs. Also, one can observe that at a given SNR and $\Gamma$, the HBF (BCD-SRCG, D-SVD) scheme outperforms HBF (BCD-SRCG, P-SVD) at a lower value of $N$ while the latter scheme approaches the former at a higher value of $N$. This is due to the fact that a large $N$ produces a higher passive beamforming gain, which results in higher IP at the PU and limits the performance of the HBF (BCD-SRCG, D-SVD) scheme. However, a large value of $N$ provides better degrees of freedom for nulling the interference, which improves the performance of the HBF (BCD-SRCG, P-SVD) scheme. 
\subsubsection{Sum-SE versus SNR}
As shown in Fig. \ref{fig:Fig3}, we compare the sum-SE of the system versus SNR for a fixed IP threshold $\Gamma=0$ dB when the number of reflective elements is $N=\{16, 32\}$. As can be seen from the figure, the sum-SE of the proposed HBF (BCD-SRCG, D-SVD) method approaches that of FBD w/o interference at low SNR and saturates at high SNR. This is because at low SNR regime, the IP constraint is inactive due to the low level of interference induced at the PU, whereas at high SNR, it becomes active due to the increased interference at the PU. Therefore, the system is limited by the quantity $\Gamma$ at high SNRs. In addition, the sum-SE of the HBF (BCD-SRCG, P-SVD) method increases with the SNR regardless of $\Gamma$, since the interference is nulled via the projection method. Also, one can observe that the HBF (BCD-SRCG, P-SVD) method outperforms the HBF (BCD-SRCG, D-SVD) scheme at high SNR due to IP limitations in the latter optimization at high SNR. Furthermore, as expected, the proposed HBF (BCD-SRCG, D-SVD) scheme has a performance edge over the naive HBF (white spectrum) scheme, which shows the effectiveness of the proportional water-filling solution toward optimal power allocation. Furthermore, it can be observed that the system having $N = 32$ reflecting elements outperforms that with $N = 16$. This trend is expected due to the higher passive beamforming gain of the former.
\begin{figure}
\centering
\includegraphics[width = 9 cm]{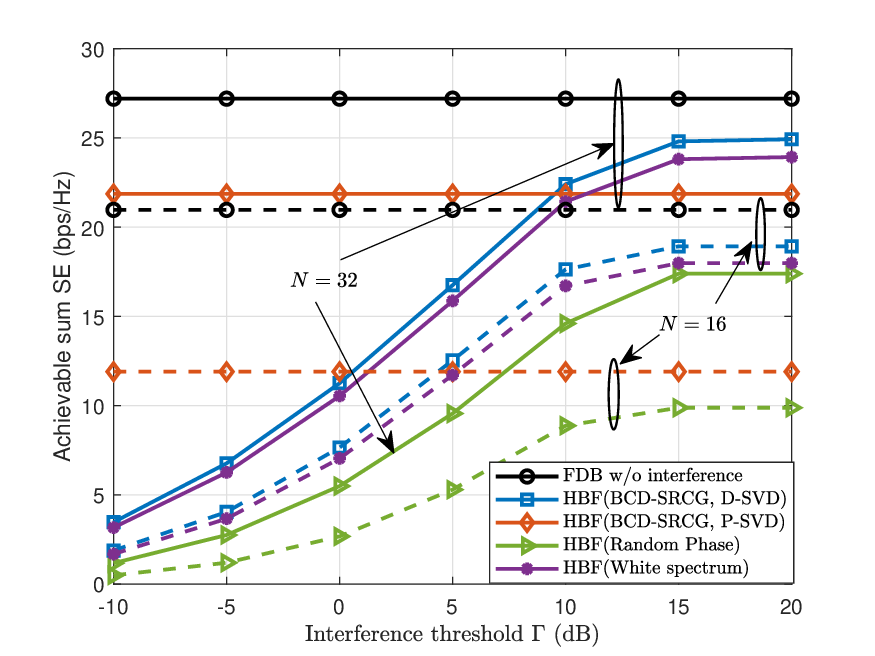}
\caption{Achievable sum-SE versus IP threshold $\Gamma$ for an $8\times128$ RIS-aided mmWave MIMO CR system with SNR$=0$ dB. All other parameters are listed in Table. \ref{tab:simulation parameters}.}
\label{fig:Fig4}
\vspace{-4mm}
\end{figure}
\begin{figure}
\centering
\includegraphics[width = 9 cm]{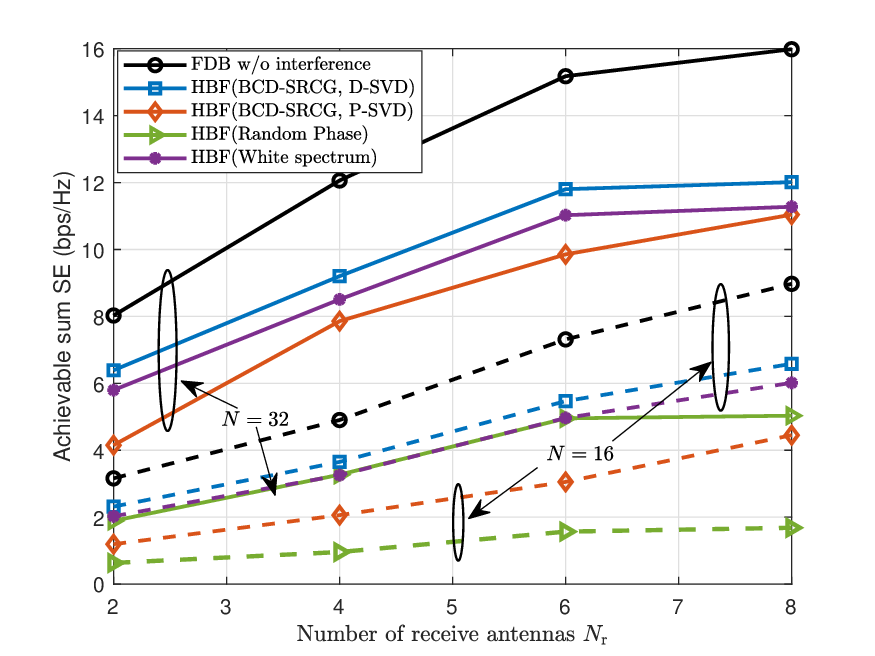}
\caption{Achievable sum-SE versus number of receive antennas $N_\mathrm{r}$ for an $N_\mathrm{r}\times128$ RIS-aided mmWave MIMO CR system with SNR$=0$ dB and $\Gamma=0$ dB. All other parameters are listed in Table. \ref{tab:simulation parameters}.}
\label{fig:Fig6}
\vspace{-4mm}
\end{figure}
\begin{figure}
\centering
\includegraphics[width = 9 cm]{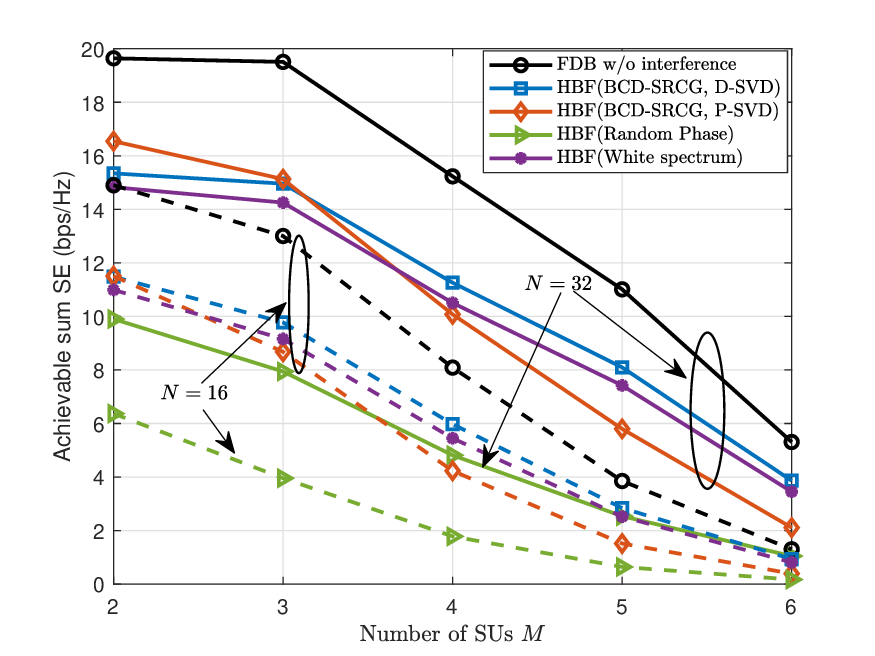}
\caption{Achievable sum-SE versus number of SUs $M$ for an $8\times128$ RIS-aided mmWave MIMO CR system with SNR$=0$ dB and $\Gamma=0$ dB. All other parameters are listed in Table. \ref{tab:simulation parameters}.}
\label{fig:Fig7}
\vspace{-4mm}
\end{figure}
\begin{figure}
\centering
\includegraphics[width = 9 cm]{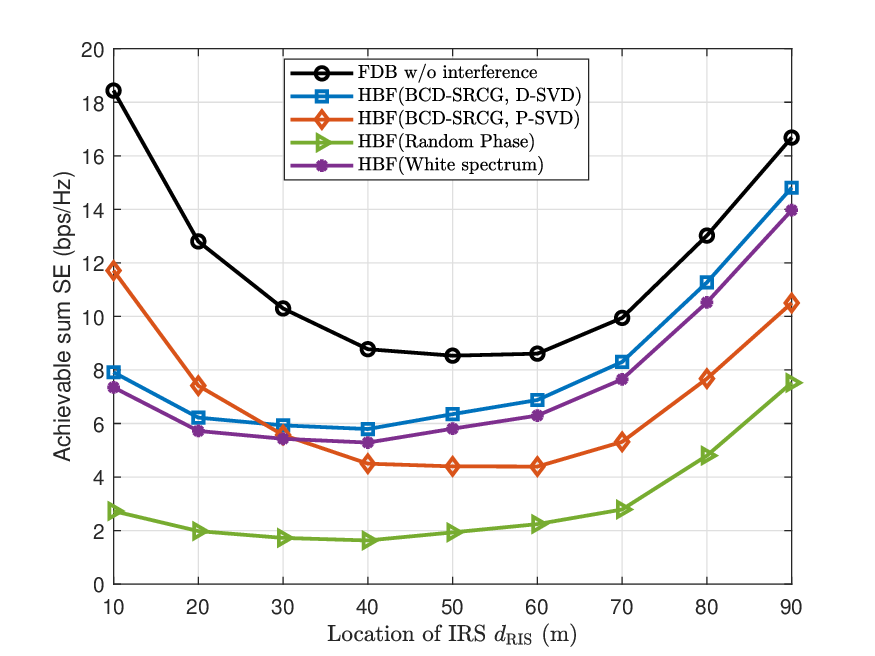}
\caption{Achievable sum-SE versus horizontal distance of RIS for an $8\times128$ RIS-aided mmWave MIMO CR system with SNR$=0$ dB and $\Gamma=0$ dB. All other parameters are listed in Table. \ref{tab:simulation parameters}.}
\label{fig:Fig8}
\vspace{-4mm}
\end{figure}

\subsubsection{Sum-SE versus $\Gamma$}
In Fig. \ref{fig:Fig4}, we plot the sum-SE of the system with respect to the IP threshold $\Gamma$ for a fixed value of $SNR = 0$ dB. It can be seen from the figure that the sum-SE of the HBF (BCD-SRCG, D-SVD) method increases with the IP threshold. This is due to the fact that the large value of $\Gamma$ provides an opportunity for the SUs to transmit at a higher power due to the improved ability of the PU to tolerate the interference. Furthermore, the sum-SEs of the HBF (BCD-SRCG, P-SVD) and the optimal w/o interference schemes are constant with respect to $\Gamma$, which shows that these schemes are independent of the IP threshold. However, note that the HBF (BCD-SRCG, D-SVD) scheme has a superior SE in comparison to the HBF (BCD-SRCG, P-SVD) in the higher $\Gamma$ regime. This is due to the fact that at sufficiently large values of $\Gamma$, the IP constraint becomes ineffective owing to the enhanced interference tolerance at the PU. Therefore, at high $\Gamma$, the system performance is only limited by the maximum value of the TP $P_\mathrm{T}$. Again, it can be seen that the higher passive beamforming gain of the $N = 32$ system makes it superior to the $N = 16$ system.
\subsubsection{Sum-SE versus $N_\mathrm{r}$}
Fig. \ref{fig:Fig6} illustrates the sum-SE of the system versus the number of RAs $N_\mathrm{r}$ for the fixed values of $SNR=0$ dB and $\Gamma=0$ dB. As expected, the sum-SE of the proposed schemes increases upon increasing $N_\mathrm{r}$ due to the increased multiplexing gain. However, observe that the sum-SE of the HBF (BCD-SRCG, D-SVD) scheme almost saturates at $N_\mathrm{r}=8$. While the performance of the HBF (BCD-SRCG, P-SVD) is poor at lower values of $N_\mathrm{r}$, it increases almost linearly as $N_\mathrm{r}$ increases, and it approaches that of the HBF (BCD-SRCG, D-SVD) at $N_\mathrm{r}=8$. This is due to the fact that a large value of $N_\mathrm{r}$ produces higher antenna gain, which increases the IP at the PU in the HBF (BCD-SRCG, D-SVD) scheme, resulting in saturation of its performance. Moreover, the HBF (BCD-SRCG, P-SVD) scheme is free of the IP constraint. As a result, its performance is not limited by the antenna gain. Moreover, observe that the performance gain of all the systems obtained by increasing the number of RAs at the SUs is significantly higher than that obtained by increasing the number of reflective elements of the RIS in Fig. \ref{fig:Fig2}. However, note that this improved performance is achieved at the cost of the high energy consumption of the former due to the increased number of active RAs.
\subsubsection{Sum-SE versus number of SUs $M$}
Furthermore, in Fig. \ref{fig:Fig7}, we plot the sum-SE of SUs vs. the number of SUs $M$ for a fixed $SNR=0$ dB and $\Gamma=0$ dB. As seen, the sum-SE of the system decreases as $M$ increases due to the increment in the MUI and reduction in the power per SU. To compensate these losses, it is advisable to increase the number of RAs in the HBF (BCD-SRCG, D-SVD) scheme as $M$ increases, but not to increase the TP, as it leads to an undesirable increase in the IP at the PU. Moreover, the HBF (BCD-SRCG, P-SVD) scheme is outperformed by the HBF (BCD-SRCG, D-SVD) scheme as $M$ increases due to loss in beamforming gain for ZF. However, it is advisable to increase the TP in the HBF (BCD-SRCG, P-SVD) scheme instead of increasing the number of RAs, because an increment in the power of this scheme does not affect the PU. Furthermore, the system performance is improved by increasing the number of reflective elements from $N=16$ to $N=32$, demonstrating that an RIS with a large number of reflective elements has an improved ability to suppress MUI.
\subsubsection{Sum-SE versus horizontal distance of the RIS}
Moreover, in Fig. \ref{fig:Fig8}, we plot the sum-SE of the system vs. the horizontal distance of RIS, denoted by $d_\mathrm{RIS}$, in the range of $10$m to $90$m, for a fixed values of $SNR=0$dB and $\Gamma=0$dB. As seen from the figure, the sum-SE of the system initially decreases as $d_\mathrm{RIS}$ increases, approaching its minimum value, and then subsequently increasing as $d_\mathrm{RIS}$ increases. Therefore, it is beneficial to place the RIS within the vicinity of the CBS or the SUs for better performance but not in the vicinity of the PU. Also, one can observe that the HBF (BCD-SRCG, P-SVD) method performs better when the RIS is closer to the CBS as the passive beamforming gain of the RIS does not affect the PU. By contrast, the HBF (BCD-SRCG, D-SVD) method performs better when the RIS is closer to the SUs because the passive gain of the RIS is affected the least when it is far from the PU and closer to the SUs.
\begin{figure}
\centering
\includegraphics[width = 8.7 cm]{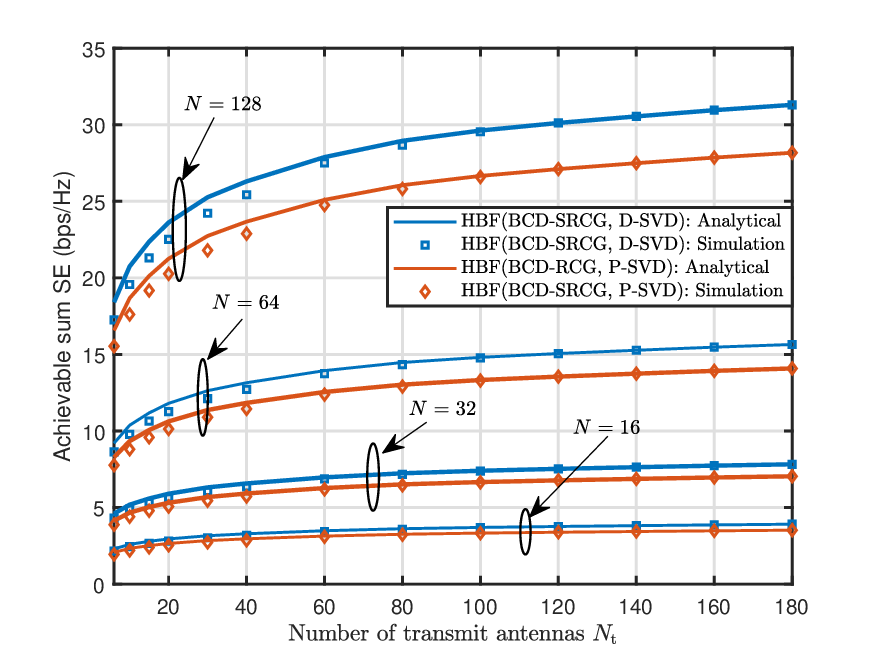}
\caption{Achievable sum-SE versus number of transmit antennas $N_\mathrm{t}$ for an $8\times N_\mathrm{t}$ RIS-aided mmWave MIMO CR system with SNR$=0$ dB and $\Gamma=0$ dB. All other parameters are listed in Table. \ref{tab:simulation parameters}.}
\label{fig:Fig9}
\vspace{-4mm}
\end{figure}
\subsubsection{Sum-SE versus $N_\mathrm{t}$}
Finally, we examine the achievable sum SE versus the number of TAs $N_\mathrm{t}$ to quantify the performance gap arising due to the assumption of near orthogonality of user channels considered in (\ref{eqn:mutual_2}) and (\ref{eqn:Y_m}), for a large number of antenna elements. Fig. \ref{fig:Fig9} plots both the analytical and Monte Carlo simulation based sum SE versus $N_\mathrm{t}$ for different numbers of reflective elements, $N=\{16, 32, 64, 128\}$ of the RIS at a fixed value of $SNR=0$ dB and $\Gamma=0$ dB. It can be seen from the figure that as $N_\mathrm{t}$ increases, the simulated values for both the HBF (BCD-SRCG, D-SVD) and HBF (BCD-SRCG, P-SVD) schemes approach the corresponding analytical values. More specifically, there are only marginal deviations of $1.38 \%, 1.26 \%, 1.21 \%$ and $ 1.03 \%$ at $N_\mathrm{t}=60$ for $N=\{16, 32, 64, 128\}$, which demonstrates the validity of the user orthogonality assumptions in the massive antenna array regime.

\section{\uppercase{Conclusion}}\label{conclusion}
We investigated the ability of the RIS technology to aid multiple SUs in a mmWave MIMO CR system operating in the underlay mode. A two-stage hybrid transceiver design was proposed based on the SRCG-BCD algorithm to jointly design the hybrid TPC/RC and RM, which maximizes the sum-SE of the secondary system, while restricting the IP induced at the PU to a predefined threshold. The proposed approach initially designs a pair of vectors for the RF TPC and RC, and each element of the RM matrix successively. Subsequently, two sub-optimal solutions were proposed to design the BB TPC/RC based on the SVD of the effective BB channel. Furthermore, the proportional water-filling approach was adopted to optimize the power allocation to each stream of each SU for the sake of user fairness. 
Finally, simulation results were presented, which show the effectiveness of the proposed schemes in RIS-aided mmWave MIMO CR systems.

\begin{appendices}
\section{\uppercase{Derivation for Eq.} (\ref{D_5})}\label{AP_C}
For the given hybrid TPC and RC obtained using the D-SVD method, one can express the achievable SE $\mathcal{R}^\mathrm{D}_m $ of the $m$th SU as
\begin{equation}
\begin{aligned} 
\mathcal{R}^\mathrm{D}_m & = \mathrm{log}_2 \Bigg (\bigg |\mathbf{I}_{N_\mathrm{s}} +  \mathbf{R}_n^{-1} \mathbf {W}^H_{\mathrm{BB},m}\mathbf{W}^H_{\mathrm{RF},m} \mathbf{H}_m \mathbf{F}_\mathrm{RF}\mathbf{F}^\mathrm{D}_{\mathrm{BB},m}  \\ 
&\qquad \times\mathcal{D}(\mathbf{p}^\mathrm{D}_m)(\mathbf{F}^\mathrm{D}_{\mathrm{BB},m})^H \mathbf{F}_\mathrm{RF}^H \mathbf{H}_m^H \mathbf{W}_{\mathrm{RF},m}\mathbf{W}_{\mathrm{BB},m} \bigg | \Bigg).
\end{aligned}
\tag{71}
\end{equation}
For a large number of antennas, one can approximate $\mathbf{W}^H_{\mathrm{BB},m} \mathbf{W}^H_{\mathrm{RF},m}\mathbf{W}_{\mathrm{RF},m}\mathbf{W}_{\mathrm{BB},m} \approx \mathbf{I}_{N_\mathrm{s}}$. Thus, $\mathcal{R}^\mathrm{D}_m $ can be approximated as
\begin{align} 
\mathcal{R}^\mathrm{D}_m & \approx \mathrm{log}_2 \Bigg (\bigg |\mathbf{I}_{N_\mathrm{s}} + \frac{1}{\sigma^2} \left(\mathbf{F}^\mathrm{D}_{\mathrm{BB},m}\right)^H \mathbf{F}_\mathrm{RF}^H \mathbf{H}_m^H \mathbf{H}_m  \nonumber\\ 
&\qquad \times \mathbf{F}_\mathrm{RF}\mathbf{F}^\mathrm{D}_{\mathrm{BB},m} \mathcal{D}(\mathbf{p}^\mathrm{D}_m)\bigg | \Bigg), \tag{72}\\
& = \mathrm{log}_2 \Bigg (\bigg |\mathbf{I}_{N_\mathrm{s}} + \frac{1}{\sigma^2} \left(\mathbf{F}^\mathrm{2,D}_{\mathrm{BB},m}\right)^H \left(\mathbf{F}^\mathrm{1,D}_{\mathrm{BB},m}\right)^H\mathbf{F}_\mathrm{RF}^H \mathbf{H}_m^H \mathbf{H}_m \nonumber \\ 
&\qquad \times \mathbf{F}_\mathrm{RF}\mathbf{F}^\mathrm{1,D}_{\mathrm{BB},m} \mathbf{F}^\mathrm{2,D}_{\mathrm{BB},m}\mathcal{D}(\mathbf{p}^\mathrm{D}_m)\bigg | \Bigg), \tag{73}\\
\mathcal{R}^\mathrm{D}_m & \overset{(c)}{=} \log_{2}\Big(\Big\vert{\bf I}_{N_{\rm s}} + \frac{1}{\sigma^2}(\mathbf{F}^\mathrm{2,D}_{{\rm BB},m})^H\widehat{\mathbf{\Sigma}}^2_m\mathbf{F}^\mathrm{2,D}_{{\rm BB},m}{\cal D}(\mathbf{p}^\mathrm{D}_m)\Big\vert\Big). \tag{74}
\end{align}
Approximation $(c)$ follows due to the fact that $\mathbf{F}_\mathrm{RF}\mathbf{F}^\mathrm{1,D}_{\mathrm{BB},m} \approx  \widehat{\mathbf{V}}_m$.
\section{PROOF OF THEOREM 1}\label{AP_B}
The equivalent convex optimization problem corresponding to $\mathcal{P}_{12}$ is given by
\begin{equation}\label{eqn:AP_B1}
\begin{aligned}
 &\mathcal{P}_{14}: \mathop{\min}\limits_{p^\mathrm{D}_{m,d}} \quad -\sum_{m=1}^{M}\sum_{d=1}^{N_{\rm s}}  \widetilde{w}^\mathrm{D}_{m,d}\log_{2}\left( 1 + \frac{\upsilon^\mathrm{D}_{m,d}\lVert \mathbf{f}^\mathrm{2,D}_{{\rm BB},m,d}\rVert_2^2}{\sigma^2} p^\mathrm{D}_{m,d}\right) \cr 
 & \qquad {\rm s.t.} \quad \text{(\ref{eqn:DSVD_1_1}), (\ref{eqn:DSVD_1_2}), (\ref{eqn:DSVD_1_3})}.
 \end{aligned}
 \tag{75}
\end{equation}
Inspired by the Karush-Kuhn-Tucker (KKT) framework, let us assume $\lambda$, $\tau^\mathrm{D}$ and $\mu^\mathrm{D}_{m,d} \forall m,d$ to be the Lagrange multipliers associated with the IP inequality, maximum TP inequality and power causality constraints in $\mathcal{P}_{14}$, respectively. Thus, the KKT conditions are given as \cite{CR_A4}
\begin{align}\label{eqn:KKT1}
-\frac{\widetilde{w}^\mathrm{D}_{m,d}\upsilon^\mathrm{D}_{m,d}\lVert \mathbf{f}^\mathrm{2,D}_{{\rm BB},m,d}\rVert_2^2}{\sigma^2 \left( 1 + \frac{\upsilon^\mathrm{D}_{m,d}\lVert \mathbf{f}^\mathrm{2,D}_{{\rm BB},m,d}\rVert_2^2}{\sigma^2} p^\mathrm{D}_{m,d}\right)} + \lambda\zeta_{m,d} + &\tau^\mathrm{D} t^\mathrm{D}_{m,d} -\mu^\mathrm{D}_{m,d} \nonumber \\
 &=0 \hspace{0.2cm}\forall m, d, \tag{76}\\
\lambda\left(I_{\rm th}-\sum_{m=1}^{M}\sum_{d=1}^{N_{\rm s}}p^\mathrm{D}_{m,d}\zeta_{m,d}\right)&=0, \tag{77}\\
\tau^\mathrm{D}\left(P_{\rm max}-\sum_{m=1}^{M}\sum_{d=1}^{N_{\rm s}}p^\mathrm{D}_{m,d}t^\mathrm{D}_{m,d}\right)&=0, \tag{78}\\
p^\mathrm{D}_{m,d}\geq 0, \mu^\mathrm{D}_{m,d}\geq 0, \mu^\mathrm{D}_{m,d}p^\mathrm{D}_{m,d}&=0 \hspace{0.2cm}\forall{m, d}. \tag{79}
\end{align}
From $(\ref{eqn:KKT1})$, the power profile can be written as 
\begin{equation}\label{eqn:water filling1}
\begin{aligned}
&p^\mathrm{D}_{m,d}\\
&=\max\left\{0, {\frac{1}{\lambda \zeta_{m,d}+\tau^\mathrm{D} t^\mathrm{D}_{m,d}}}-\frac{\sigma^2}{\widetilde{w}^\mathrm{D}_{m,d}\upsilon^\mathrm{D}_{m,d}\lVert \mathbf{f}^\mathrm{2,D}_{{\rm BB},m,d}\rVert_2^2}\right\} \forall m, d.
 \end{aligned}
 \tag{80}
\end{equation}
Note that the quantities $\lambda$ and $\tau^\mathrm{D}$ in $(\ref{eqn:water filling1})$ can be found using the interior point method so that the KKT conditions are satisfied.
\end{appendices}
\bibliographystyle{IEEEtran}
\bibliography{biblio.bib}

\end{document}